\documentclass[useAMS,usenatbib]{mn2e}

\usepackage{amssymb}
\usepackage{times}
\usepackage{graphicx}
\usepackage{epstopdf}
\usepackage{subfigure}
\usepackage[T1]{fontenc}
\usepackage{aecompl} 
\usepackage{multirow}
\usepackage{pdflscape}
\usepackage{rotating}

\usepackage[fleqn]{amsmath}
\usepackage{mathrsfs}
\usepackage{xcolor}

\usepackage{anyfontsize}

\usepackage{hyperref}

\newcommand{\beq}{\begin{equation}}
\newcommand{\eeq}{\end{equation}}

\def\gs{\mathrel{\lower0.6ex\hbox{$\buildrel {\textstyle >}\over{\scriptstyle \sim}$}}}
\def\ls{\mathrel{\lower0.6ex\hbox{$\buildrel {\textstyle <}\over{\scriptstyle \sim}$}}}
\newcommand{\simgt}{\lower.5ex\hbox{$\; \buildrel > \over \sim \;$}}
\newcommand{\simlt}{\lower.5ex\hbox{$\; \buildrel < \over \sim \;$}}

\newcommand{\aap}{A\&A}
\newcommand{\apj}{ApJ}
\newcommand{\apjl}{ApJ}
\newcommand{\apjs}{ApJS}
\newcommand{\aj}{AJ}

\newcommand{\pasj}{PASJ}

\newcommand{\mnras}{MNRAS}

\newcommand{\ssr}{Space Science Reviews}

\defcitealias{se+et14_comalit_I}{CoMaLit-I}
\defcitealias{ser+al14_comalit_II}{CoMaLit-II}
\defcitealias{ser14_comalit_III}{CoMaLit-III}
\defcitealias{se+et14_comalit_IV}{CoMaLit-IV}

\begin{document}

\title[$Y_{500}$-$M_{500}$ scaling relation]{CoMaLit -- II. The scaling relation between mass and Sunyaev-Zel'dovich signal for Planck selected galaxy clusters}
\author[M. Sereno, S. Ettori, L. Moscardini]{
Mauro Sereno$^{1,2}$\thanks{E-mail: mauro.sereno@unibo.it (MS)}, Stefano Ettori$^{2,3}$, Lauro Moscardini$^{1,2,3}$
\\
$^1$Dipartimento di Fisica e Astronomia, Alma Mater Studiorum -- Universit\`a di Bologna, Viale Berti Pichat 6/2, 40127 Bologna, Italia\\
$^2$INAF, Osservatorio Astronomico di Bologna, via Ranzani 1, 40127 Bologna, Italia\\
$^3$INFN, Sezione di Bologna, viale Berti Pichat 6/2, 40127 Bologna, Italia\\
}


\maketitle

\begin{abstract}
We discuss the scaling relation between mass and integrated Compton parameter of a sample of galaxy clusters from the all-sky {\it Planck} Sunyaev-Zel'dovich catalogue. Masses were measured with either weak lensing, caustics techniques, or assuming hydrostatic equilibrium. The retrieved $Y_{500}$-$M_{500}$ relation does not strongly depend on the calibration sample. We found a slope of 1.4-1.9, in agreement with self-similar predictions, with an intrinsic scatter of $20\pm10$ per cent. The absolute calibration of the relation can not be ascertained due to systematic differences of $\sim$20-40 per cent in mass estimates reported by distinct groups. Due to the scatter, the slope of the conditional scaling relation, to be used in cosmological studies of number counts, is shallower, $\sim$1.1-1.6. The regression methods employed account for intrinsic scatter in the mass measurements too. We found that Planck mass estimates suffer from a mass dependent bias.
\end{abstract}


\section{Introduction}

Clusters of galaxies are big and interesting \citep{voi05,lim+al13}. Theory and numerical simulations of their formation and evolution usually characterise clusters by their mass. Observations at a variety of wave-lengths can probe the X-ray luminosity and temperature, the optical richness and luminosity, the Sunyaev-Zel'dovich (SZ) flux, the weak lensing (WL) shear and convergence, and so on. 

A crucial piece of the puzzle is the accurate knowledge of scaling relations between cluster properties. Scaling relations enable us to confront theoretical predictions with actual data and test at once cosmological models \citep{planck_2013_XX} and cluster physics \citep{bat+al12,ett13}. Mass-observable scaling relations are at the basis of all work that exploits the abundance of galaxy clusters for constraining cosmological parameters \citep{vik+al09,man+al10,roz+al10}.

The mass dependence of cluster observables is strictly connected to the main gravitational processes driving the cluster evolution \citep{kai86,gio+al13}. Departures from these self-similar expectations show imprints of the non gravitational phenomena which occur during the formation and evolution of clusters, such as feedback and non-thermal processes \citep{mau+al12,ett13}

Likewise, the scatter around the scaling relations probes the variety of cluster radial structures and morphologies, the state of the intracluster gas, the presence or absence of a cool core, and the dynamical state \citep{arn+al10,ras+al12,ett15}.

The ability to robustly measure empirical cluster scaling relations is crucial \citep{roz+al14b}. In this context, the relation between mass and SZ flux has a prominent role. It is expected to have small intrinsic scatter \citep{kay+al12,bat+al12} and must be accurately calibrated to constrain cosmological parameters using number counts \citep{planck_2013_XX}. As it is usual for scaling relations, it is modelled as a power law,
\beq
\label{eq_scal_rela}
\left[\frac{H(z)}{H_0} \right]^{-2/3}\left[ \frac{D_\mathrm{A}^2(z)Y_{500}}{10^{-4}\mathrm{Mpc}^2}\right]=10^{\alpha}\left[ \frac{M_{500}}{M_\mathrm{pivot}}\right]^\beta,
\eeq
where $H(z)$ is the redshift dependent Hubble parameter, $H_0\equiv H(z=0)$, $D_\mathrm{A}$ is the angular diameter distance to the cluster, $M_\mathrm{pivot}$ is a pivotal mass, and $M_{500}$ and $Y_{500}$ denote the mass and the spherically integrated Compton parameter within a sphere of radius $r_{500}$, which encloses a mean over-density of 500 times the critical density at the cluster redshift.

Planck's cluster count cosmology results \citep{planck_2013_XX} differ from those from the Planck's measurements of the primary CMB (Cosmic Microwave Background) temperature anisotropies \citep{planck_2013_XVI}. Discrepancies may hinge on inaccurate calibration of the scaling relation. The larger values of the amplitude of the matter power spectrum, $\sigma_8$, and of the matter density parameter, $\Omega_\mathrm{M}$, preferred by CMB experiments can be accommodated by a mass bias of about 45 per cent, where the bias is defined  through the ratio of the measured to the  `true'  mass $M_{500}^\mathrm{Meas}=(1-b)M_{500}$ \citep{planck_2013_XX}. This level of bias is larger than expected. Based on numerical simulations, \citet{planck_2013_XX} adopted a default value of $b=0.2$, which was assumed as a prior in their analysis. From comparison with a sample of weak lensing masses, \citet{lin+al14} found a larger bias,  $b =0.30\pm0.06$.

Self-similar models predict that the slope of the $Y_{500}$-$M_{500}$ relation (in logarithmic variables) is $\beta=5/3$ \citep{kai86,gio+al13,ett13}. Self-similarity only accounts for gravity processes, which are dominant at the scales of the massive clusters selected by SZ flux. Numerical simulations confirm that the $Y_{500}$-$M_{500}$ relation has little scatter ($\ls 10$ per cent in the local Universe), it is nearly insensitive to the cluster gas physics and it evolves to redshift $\ls 1$ in agreement with the self-similar expectation of $\beta \sim 1.6-1.8$ \citep{sta+al10,bat+al12,kay+al12}.

This is the second of a series of papers devoted to the COmparison of galaxy cluster MAsses in LITerature (CoMaLit) and to the calibration of scaling relations. In the first paper \citep[ henceforth CoMaLit-I]{se+et14_comalit_I}, we compared two well regarded mass proxies, the weak lensing (WL) mass and the X-ray determination of the mass based on the hypothesis of hydrostatic equilibrium (HE). We measured the intrinsic scatters, estimated the relative bias, and discussed how the intrinsic scatter in the mass proxy affects the scaling relations. In this second paper, we develop and apply the formalism of \citetalias{se+et14_comalit_I} to calibrate the SZ flux estimated by the Planck satellite against mass estimates. We considered masses obtained either from WL analyses \citep{wtg_III_14,hoe+al12,ume+al11b}, X-ray studies \citep{don+al14,lan+al13}, or the caustic technique \citep[][ CS]{ri+di06}. The third paper of the series \citep[ CoMaLit-III]{ser14_comalit_III} introduces the Literature Catalogs of weak Lensing Clusters of galaxies (LC$^2$),  which are standardised compilations of clusters with measured WL masses. In the fourth paper of the series \citep[ CoMaLit-IV]{se+et14_comalit_IV}, the Bayesian methodology is extended to account for time-evolution of the scaling relation. Material concerning the CoMaLit series, and future updates, will be hosted at \url{http://pico.bo.astro.it/\textasciitilde sereno/CoMaLit/}.

The paper is structured as follows. Section~\ref{sec_sz_samp} reviews some properties of the Planck selected clusters. Section~\ref{sec_samp} lists the cluster samples used to calibrate the scaling relation. The mass bias affecting Planck masses is discussed in Sec.~\ref{sec_mass_bias}. Section~\ref{sec_cali_samp} is devoted to highlight some features of the Planck calibration sample. The regression scheme used to derive the scaling relations is discussed in Sec.~\ref{sec_scal_rela}. Results are presented in Sec.~\ref{sec_resu}. Discussion of results and final considerations are contained in Secs.~\ref{sec_disc}, and ~\ref{sec_conc}, respectively. Appendix~\ref{app_slop} discusses the meaning of the slope of a scaling relation.  

Conventions and notations are as in \citetalias{se+et14_comalit_I}. We assumed a flat $\Lambda$CDM cosmology with density parameter $\Omega_\mathrm{M}=0.3$, and Hubble constant $H_0=70~\mathrm{km~s}^{-1}\mathrm{Mpc}^{-1}$; `$\log$' is the logarithm to base 10, and `$\ln$' is the natural logarithm.

\section{SZ sample}
\label{sec_sz_samp}

As reference sample of SZ clusters, we considered the clusters from the first Planck SZ Catalogue \citep[PSZ1,][]{planck_2013_XXIX} detected with the Matched Multi-filter method MMF3. This algorithm discovered 883 candidates with a signal to noise ratio (SNR) above 4.5 outside the highest-emitting Galactic regions, the Small and Large Magellanic Clouds, and the point source masks.\footnote{We used the `COM\_PCCS\_SZ-validation\_R1.13.fits' catalog and the additional information in the `COM\_PCCS\_SZ-union\_R1.12.fits' catalog, which are available from the Planck Legacy Archive at \url{http://pla.esac.esa.int/pla/aio/planckProducts.html}.} 

The redshift determination is available for 664 clusters. The PSZ1 catalogue spans a broad mass range from $0.1$ to $16\times10^{14}M_\odot$ at a median redshift of $z\sim 0.22$. 

A purer subsample of 189 candidates constructed by selecting the detections above a SNR threshold of 7 over 65 per cent of the sky constitutes the cosmological sample \citep{planck_2013_XX}. 

 \citet{planck_2013_XX} determined the $Y_{500}$-$M_{500}$ relation through multiple steps. Firstly, the local $Y_\mathrm{X}$-$M^\mathrm{HE}_{500}$ relation, where $Y_\mathrm{X}$ is the X-ray analogue of $Y_{500}$ and it is defined as the product of the gas mass within $r_{500}$ and the spectroscopic temperature outside the core \citep{kra+al06}, was calibrated against the hydrostatic masses $M_{500}^\mathrm{HE}$ of 20 relatively relaxed local clusters \citep{arn+al10}. This local sample is not representative of the SZ selected Planck clusters. Masses estimated through this scaling relation are denoted $M_{500}^{Y_\mathrm{X}}$. 

Secondly, the $Y_{500}$-$M_{500}^{Y_\mathrm{X}}$ relation was computed for 71 detections from the Planck cosmological sample for which good quality XMM-Newton observations were available. The SZ signals were re-estimated within a sphere of radius $r^{Y_\mathrm{X}}_{500}$, centred on the position of the X-ray peak. Standard redshift evolution was assumed.
 
For the scaling $Y_{500}$-$M_{500}^{Y_\mathrm{X}}$, \citet{planck_2013_XX} found $\alpha=-0.186 \pm0.011$ and $\beta=1.79\pm0.06$ for $M_\mathrm{pivot}=6\times10^{14}M_\odot$, see Eq.~(\ref{eq_scal_rela}). The Planck team argued that a bias $b$ can still persist in the HE mass measurements, $M_{500}^\mathrm{HE}=(1-b)M_{500}$. Based on a suite of numerical simulations \citep{bat+al12,kay+al12}, they estimated $b=0.2^{+0.1}_{-0.2}$. 

The scaling relation was then used to break the size-flux degeneracy and to estimate cluster masses. In what follows, $M_{500}^{Y_z}$ denotes the mass estimated by the Planck team through the scaling relation $Y_{500}$-$M_{500}^{Y_\mathrm{X}}$.

\section{Calibration samples}
\label{sec_samp}

\begin{table*}
\caption{Calibration samples. Col.~1: name. Col.~2: type of mass measurement (`WL' for weak-lensing analyses; `CS' for the caustic technique; `HE' for hydrostatic masses). Col.~3: type of sample (`F' for the full sample of PSZ1 detections; `C' for PSZ1 clusters in the cosmological subsample). Col.~4: number of clusters in the sample detected by Planck-MMF3. Cols.~5 and 6: typical redshift and dispersion. Cols.~7 and 8: typical mass and dispersion. Col.~9: main references. Typical values and dispersions are computed as bi-weighted estimators. Masses are in units of $10^{14}M_\odot$.}
\label{tab_samples}
\centering
\begin{tabular}[c]{lccrllrcl}
	\hline
	Name			&	Mass	&	Sample	&	$N_\mathrm{Cl}$	&	$z$	&	$\sigma_z$	&	$M_{500}$	&	$\sigma_{M_{500}}$  & References\\
	\hline
	LC$^2$-{\it single}	&	WL	&	F	&	115	&	0.24	&	0.15	&	7.3	&	5.2	&	\citet{ser14_comalit_III}	\\
	LC$^2$-{\it single}	&	WL	&	C	&	65	&	0.22	&	0.13	&	8.9	&	5.7	&	\citet{ser14_comalit_III}	\\
	WTG			&	WL	&	F	&	34	&	0.35	&	0.13	&	11.8	&	5.2	&	\citet{wtg_III_14}	\\
	WTG			&	WL	&	C	&	22	&	0.31	&	0.13	&	12.4	&	5.6	&	\citet{wtg_III_14}	\\
	CLASH-WL		&	WL	&	F	&	11	&	0.40	&	0.13	&	11.0	&	3.7	&	\citet{ume+al14}	\\
	CLASH-WL		&	WL	&	C	&	6	&	0.37	&	0.13	&	13.7	&	5.4	&	\citet{ume+al14}	\\
	CCCP-WL			&	WL	&	F	&	35	&	0.22	&	0.07	&	7.2	&	3.0	&	\citet{hoe+al12,mah+al13}	\\
	CCCP-WL			&	WL	&	C	&	19	&	0.21	&	0.04	&	8.2	&	3.2	&	\citet{hoe+al12,mah+al13}	\\
	\hline
	CIRS			&	CS	&	F	&	22	&	0.08	&	0.02	&	1.9	&	1.3	&	\citet{ri+di06}	\\
	CIRS			&	CS	&	C	&	10	&	0.08	&	0.02	&	2.9	&	2.4	&	\citet{ri+di06}	\\
	\hline
	E10				&	HE	&	F	&	34	&	0.19	&	0.07	&	7.1	&	3.1	&	\citet{ett+al10}	\\
	E10				&	HE	&	C	&	27	&	0.20	&	0.06	&	7.3	&	3.2	&	\citet{ett+al10}	\\
	CLASH-CXO		&	HE	&	F	&	12	&	0.39	&	0.13	&	10.4	&	7.6	&	\citet{don+al14}	\\
	CLASH-CXO		&	HE	&	C	&	7	&	0.35	&	0.13	&	13.2	&	6.3	&	\citet{don+al14}	\\
	L13				&	HE	&	F	&	29	&	0.23	&	0.05	&	6.1	&	2.6	&	\citet{lan+al13}	\\
	L13				&	HE	&	C	&	21	&	0.22	&	0.05	&	6.6	&	2.4	&	\citet{lan+al13}	\\
	CCCP-HE			&	HE	&	F	&	33	&	0.22	&	0.08	&	7.1	&	3.8	&	\citet{mah+al13}	\\
	CCCP-HE			&	HE	&	C	&	19	&	0.21	&	0.04	&	7.8	&	3.7	&	\citet{mah+al13}	\\
	\hline
	\end{tabular}
\end{table*}

\begin{figure}
       \resizebox{\hsize}{!}{\includegraphics{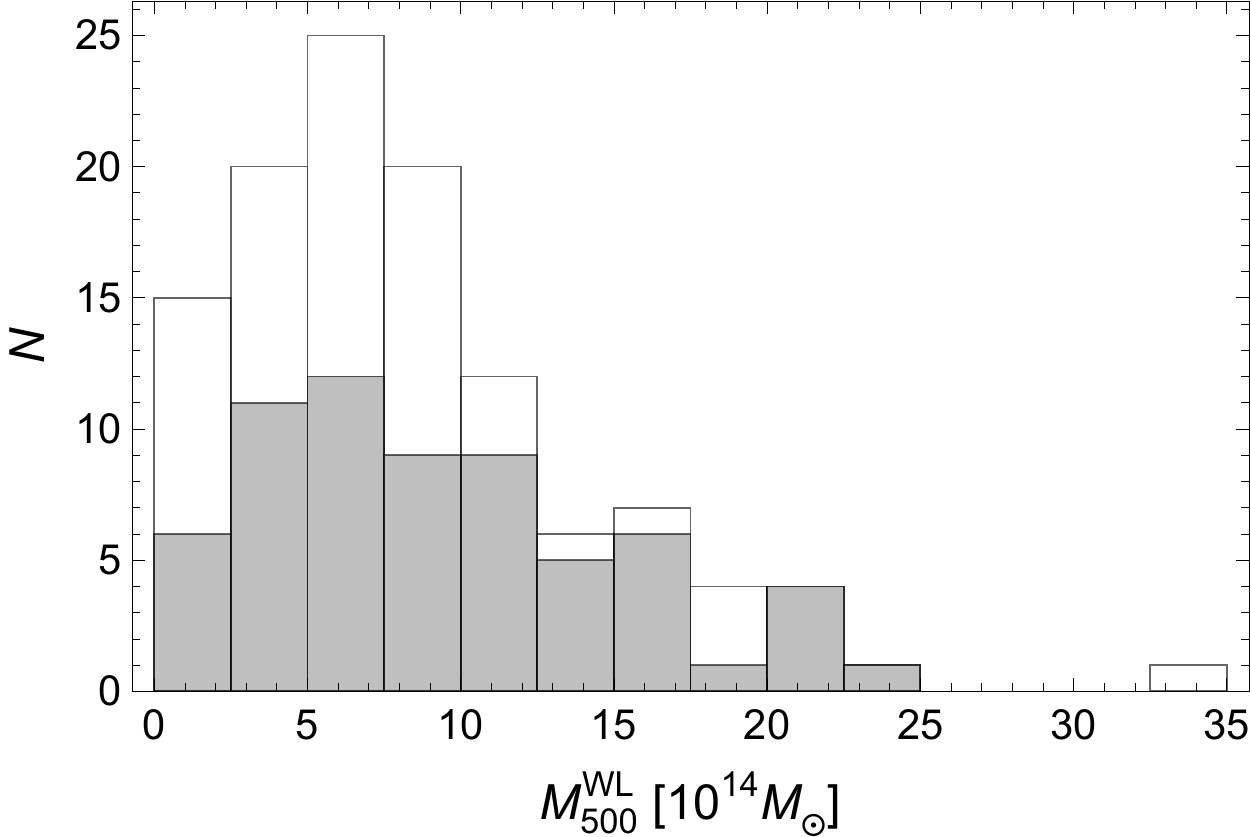}}
       \caption{Distribution of the masses $M_{500}^\mathrm{WL}$ of the clusters in the LC$^2$-{\it single} with Planck detections. The white (grey) histogram represents the full (cosmological) sample.}
	\label{fig_histo_M500_all}
\end{figure}

Up to date, there is no statistically complete and copious sample of galaxy clusters with rigorous selection criteria, substantial overlap with the PSZ1 catalog, and direct mass determinations, i.e., mass determinations not relying on scaling relations. 

We then considered a number of samples with either WL, CS, or HE masses. The main properties of the samples are listed in Table~\ref{tab_samples}. Most of the samples were introduced in \citetalias{se+et14_comalit_I}, which we refer to for details and complete references. 

The list of samples in Table~\ref{tab_samples} sightly differs from the ensemble of catalogs considered in \citetalias{se+et14_comalit_I}. Due to the small number of clusters, we did not consider HE mass determinations in the CLASH sample based on XMM observations. Furthermore, we did not consider the X-ray sample in \citet{bon+al12}, since their mass determinations exploited SZ data from the Sunyaev-Zel'dovich array, which might cause tension with the SZ determination from Planck.

On the other hand, in addition to the samples introduced in \citetalias{se+et14_comalit_I}, we also considered a sample of masses estimated with the caustic technique \citep{ri+di06}, and the LC$^2$-{\it single}, which we will briefly describe in the following.

The samples span a large range of masses and redshifts. CLASH and WTG clusters are very massive and at the larger redshifts. The CIRS clusters, which are quite small and near, lie at the other end of the spectrum. The overlap between the CLASH and the PSZ1 samples is quite small. Nevertheless we considered it because it enabled us to perform checks at the very massive end of the cluster distribution. The sub-samples of cosmological clusters differentiate themselves from the parent samples neither by mass nor by redshift.

\subsection{LC$^2$-{\it single}}

\citet{ser14_comalit_III} collected from literature data for 822 groups and clusters of galaxies with estimated redshift and WL mass. The LC$^2$-{\it single} contains 485 unique entries in the redshift range $0.02\ls z\ls 1.5$. Values of $M_{500}$ were either directly taken from the original papers or extrapolated using the quoted density profiles. If necessary, mass estimates were rescaled to the reference cosmological model. 

The clusters of this sample were discovered in a variety of ways within X-ray, optical, SZ, or WL surveys. The ensemble is very heterogeneous either for the observational facilities, the data analysis, or the selection criteria. As a consequence, the sample is not statistical, which might bias the results. A well-defined selection function can not be implemented and sample inhomogeneity might increase the observed scatter. 

On the positive side, the different finding techniques can average out some systematic biases which affect particular subsamples. Biases due to the orientation and internal structure of clusters and the projection effect of large-scale structure are strongly mitigated for a large sample. Since the only criterion for selection is the identification by Planck and because of the variety of finding techniques, we do not expect that the sample suffers from biases plaguing lensing selected samples, such as the over-concentration problem and the orientation bias \citep{og+bl09,men+al11,se+zi12}.

Very massive clusters at intermediate redshifts are preferential targets for both WL and SZ analyses. 115 WL clusters were identified by Planck too. 65 of them are included in the cosmological sample. The mass distribution is plotted in Fig.~\ref{fig_histo_M500_all}. The distribution of the cosmological clusters is similar to the full population. According to a Kolmogorov-Smirnov test, there is a 57.5 per cent probability that the masses of the cosmological and of full samples are drawn from the same distribution. The $z$ distributions are compatible at the 52.3 per cent level.

As a matter of fact, WL clusters effectively sample the massive end of the PSZ1 catalog. This can be verified quantitatively with a Kolmogorov-Smirnov test comparing the distributions of SNR. For detections with SNR above 7 (10), there is a 13.5 (91.0) per cent probability that the 68 (35) PSZ1 clusters with WL measurements out of the total 232 (86) SZ detected clusters follow the same distribution.

\subsection{Cluster Infall Regions in SDSS}

The Cluster Infall Regions in SDSS program \citep[CIRS,][]{ri+di06} studied a sample of 72 nearby clusters from the Data Release 4 of the Sloan Digital Sky Survey \citep[SDSS,][]{ade+al06} after selection in X-ray flux and redshift ($z<0.1$). 

Masses were derived from the infall patterns with the caustic technique. We computed $M_{500}$ from the published values of $r_{500}$ and $z$. Errors were rescaled from mass uncertainties at larger radii.

Only nine clusters from the CIRS sample have a WL mass estimate. Unfortunately, WL mass determinations for low redshift clusters are usually very noisy. For the CIRS clusters with WL mass, we found a mass ratio $M_{500}^\mathrm{CS}/M_{500}^\mathrm{WL}$ with central value $0.93$ and a very large scatter of 1.10.

\section{Mass bias}
\label{sec_mass_bias}

\begin{figure*}
\begin{tabular}{cc}
\includegraphics[width=8.5cm]{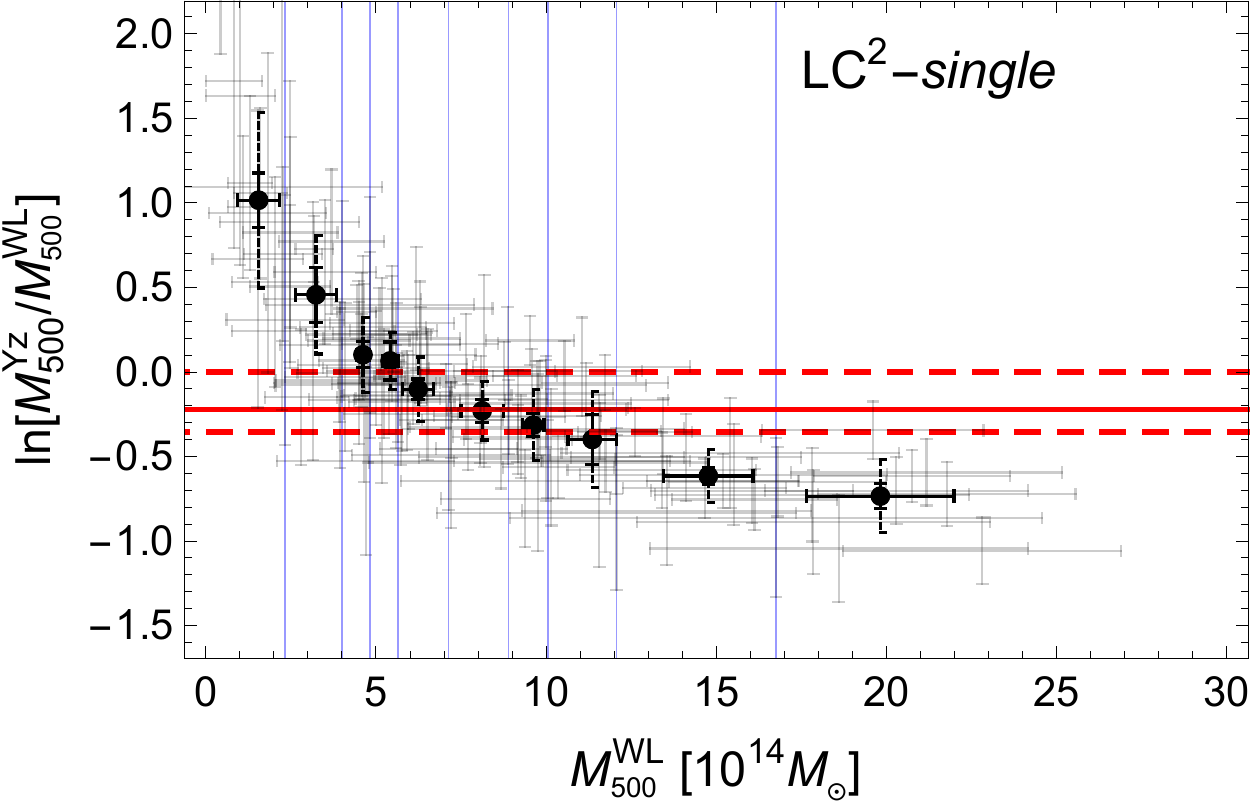} &\includegraphics[width=8.5cm]{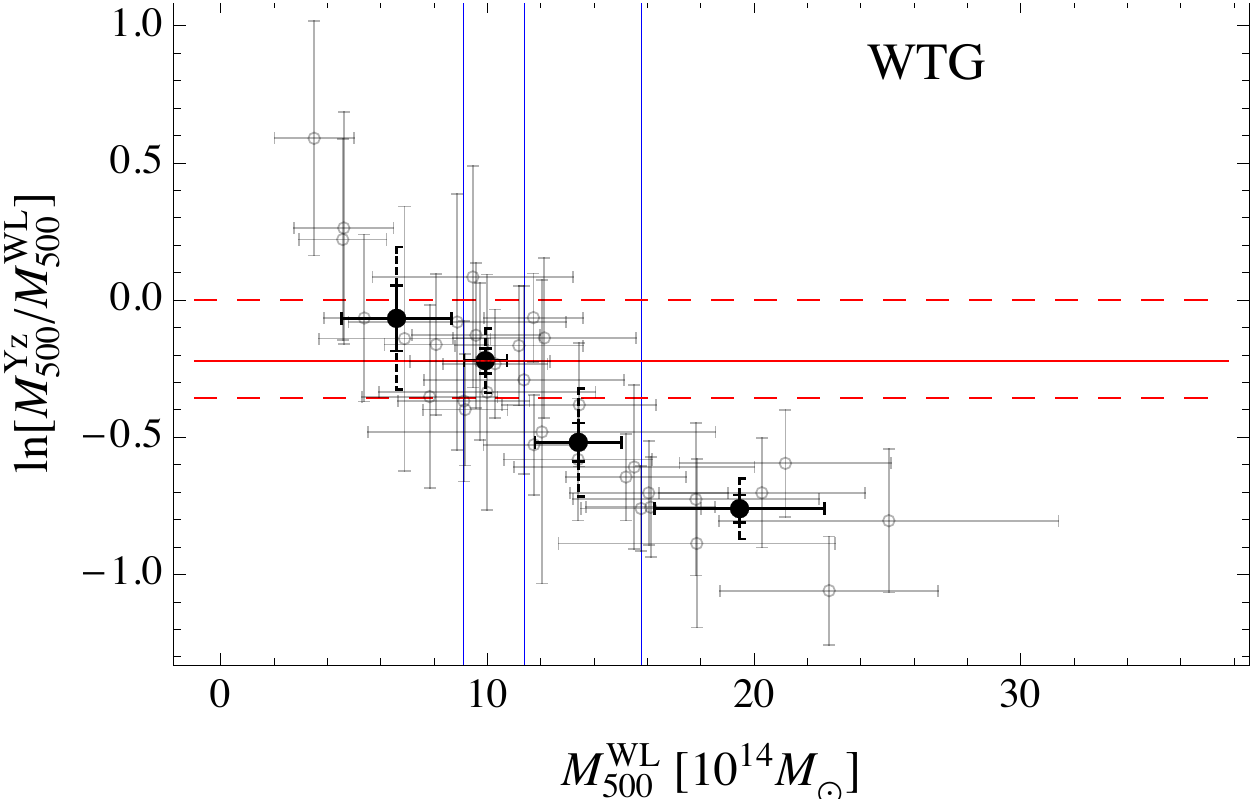}\\
\noalign{\smallskip}  
\includegraphics[width=8.5cm]{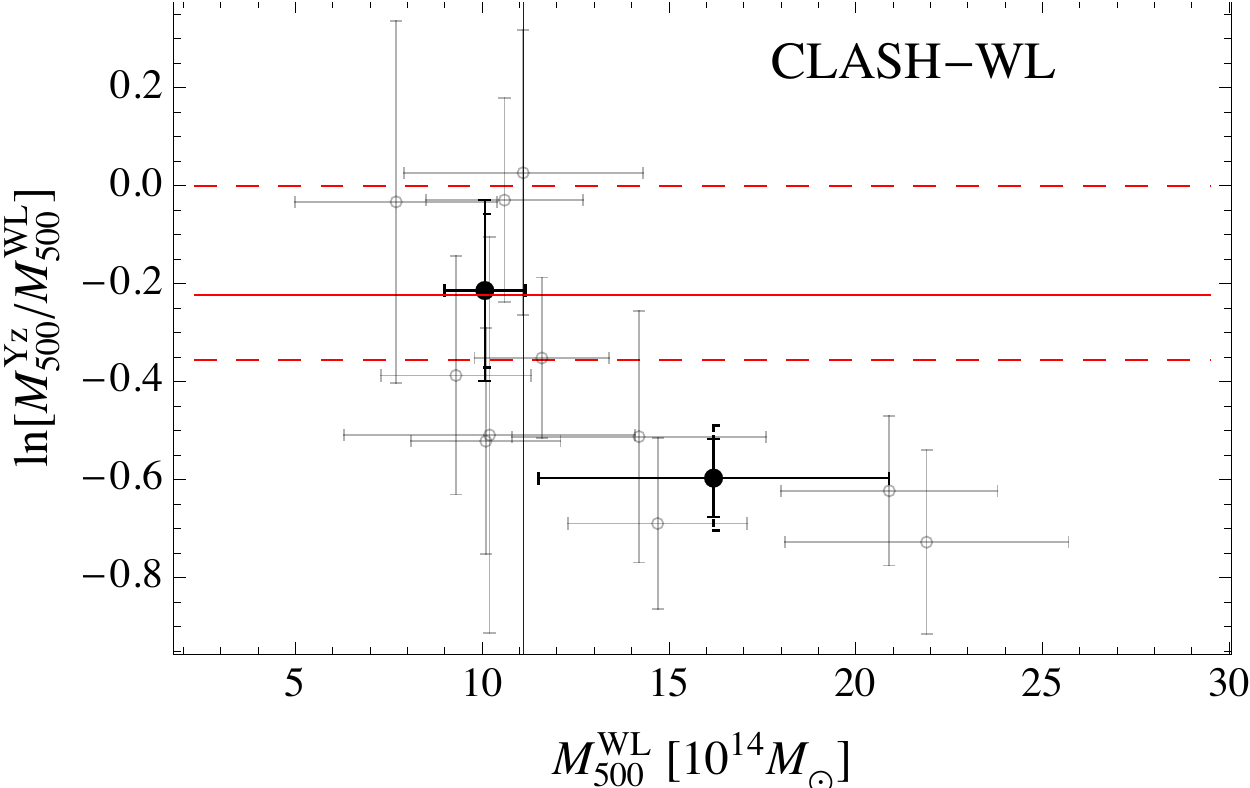} & \includegraphics[width=8.6cm]{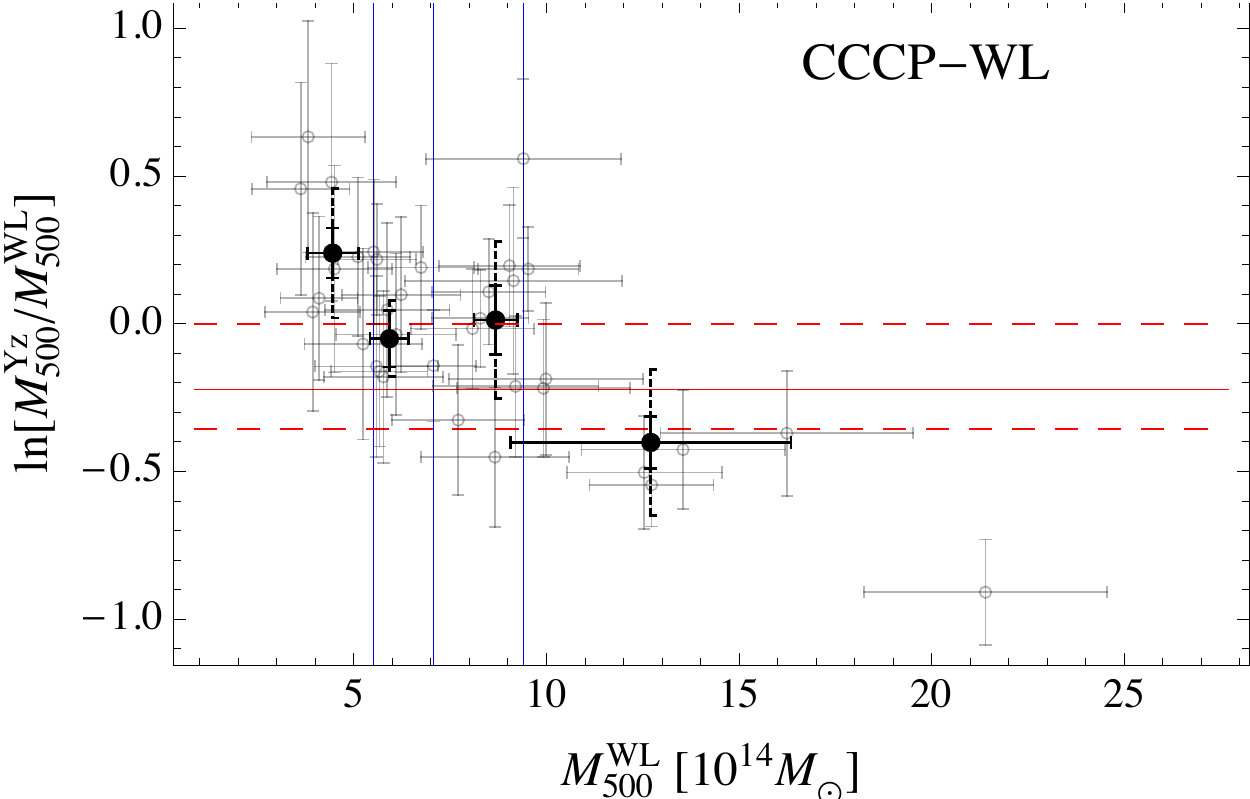}
\end{tabular}
\caption{Mass comparison, $\ln (M^{Y_z}_{500}/M^\mathrm{WL}_{500})$, for Planck clusters with WL masses, as a function of $M^\mathrm{WL}_{500}$. Thick black points mark the values for clusters binned according to their measured WL mass. Vertical blue lines mark the boundaries of the binning regions. The full error-bars for the binned points denote the 1-$\sigma$ uncertainties for the central estimate. The dashed error-bars denote the dispersion. Red lines represents the bias and the relative 1-$\sigma$ confidence region estimated by \citet{planck_2013_XX}, i.e., $b^{Y_z}=0.2^{+0.1}_{-0.2}$. The top left, top right, bottom left, and bottom right panels show the results for the LC$^2$-{\it single}, the WTG,  the CLASH-WL, and the CCCP-WL samples, respectively.}
\label{fig_bias_M500_WL}
\end{figure*}

\begin{table}
\caption{Mass comparison, $\ln M_{500}^{Y_z} -\ln M_{500}^\mathrm{WL} =\ln (1-b^{Y_z})$, for Planck clusters with WL masses. Col.~1: mass catalog used for the calibration. Col.~2: subsample (`F' stands for the full sample of Planck clusters with WL analyses; `C' refers to the Planck clusters in the cosmological subsample; `R' refers to the relaxed clusters ). Col.~3: number of clusters, $N_\mathrm{Cl}$. Cols.~4 and 5: central estimate $\mu =\langle \ln (1-b^{Y_z}) \rangle$ and scatter $\sigma$. $\mu$ and $\sigma$ are computed as bi-weighted estimators.}
\label{tab_bias_M500_WL}
\centering
\begin{tabular}[c]{l  c r r@{$\,\pm\,$}l  r@{$\,\pm\,$}l}
	\hline
	Calibration		&	Sample	&	$N_\mathrm{Cl}$	&	\multicolumn{2}{c}{$\mu$}	&	\multicolumn{2}{c}{$\sigma$}  \\
	\hline
	LC$^2$-{\it single}	&	F	&	115	&	-0.14	&	0.05	&	0.53	&	0.05		\\
	LC$^2$-{\it single}	&	F	&	64	&	-0.14	&	0.06	&	0.48	&	0.05		\\
	\hline
	WTG		&	F	&	34	&	-0.37	&	0.07	&	0.36	&	0.05		\\
	WTG		&	C	&	22	&	-0.37	&	0.11	&	0.38	&	0.06		\\
	\hline
	CLASH-WL	&	F	&	11	&	-0.45	&	0.10	&	0.29	&	0.07		\\
	CLASH-WL	&	C	&	6	&	-0.43	&	0.22	&	0.34	&	0.15		\\
	\hline
	CCCP-WL		&	F	&	35	&	0.00		&	0.06	&	0.33	&	0.05		\\
	CCCP-WL		&	C	&	19	&	0.10		&	0.09	&	0.34	&	0.09		\\
	\hline
	\end{tabular}
\end{table}

\begin{table}
\caption{Mass comparison, $\ln M_{500}^{Y_z} -\ln M_{500}^\mathrm{CS} =\ln (1-b^{Y_z})$, for Planck clusters with caustic masses. Columns are as in Table~\ref{tab_bias_M500_WL}.}
\label{tab_bias_M500_CS}
\centering
\begin{tabular}[c]{l  c r r@{$\,\pm\,$}l  r@{$\,\pm\,$}l}
	\hline
	Calibration		&	Sample	&	$N_\mathrm{Cl}$	&	\multicolumn{2}{c}{$\mu$}	&	\multicolumn{2}{c}{$\sigma$}  \\	\hline
	CIRS		&	F	&	22	&	0.35		&	0.18	&	0.79	&	0.13		\\
	CIRS		&	C	&	10	&	0.43		&	0.30	&	0.76	&	0.16		\\
	\hline
	\end{tabular}
\end{table}

\begin{table}
\caption{Mass ratio, $\ln M_{500}^{Y_z} -\ln M_{500}^\mathrm{HE} =\ln (1-b^{Y_z})$, for Planck clusters with HE masses. Columns are as in Table~\ref{tab_bias_M500_WL}.}
\label{tab_bias_M500_HE}
\centering
\begin{tabular}[c]{l  c r r@{$\,\pm\,$}l  r@{$\,\pm\,$}l}
	\hline
	Calibration		&	SZ sample	&	$N_\mathrm{Cl}$	&	\multicolumn{2}{c}{$\mu$}	&	\multicolumn{2}{c}{$\sigma$}  \\	\hline
	E10			&	F	&	34	&	-0.14	&	0.08	&	0.39	&	0.04		\\
	E10			&	C	&	27	&	-0.05	&	0.09	&	0.39	&	0.05		\\
	E10			&	R	&	10	&	-0.39	&	0.07	&	0.19	&	0.07		\\
	\hline
	CLASH-CXO	&	F	&	12	&	-0.18	&	0.16	&	0.45	&	0.11		\\
	CLASH-CXO	&	C	&	7	&	-0.24	&	0.21	&	0.37	&	0.16		\\
	CLASH-CXO	&	R	&	3	&	\multicolumn{2}{c}{$\sim-0.24$}	&	\multicolumn{2}{c}{}		\\
	\hline
	L13			&	F	&	29	&	0.21		&	0.07	&	0.34	&	0.07		\\
	L13			&	C	&	21	&	0.20		&	0.07	&	0.29	&	0.06		\\
	\hline
	CCCP-HE		&	F	&	33	&	0.07		&	0.07	&	0.43	&	0.06		\\
	CCCP-HE		&	C	&	19	&	0.13		&	0.13	&	0.42	&	0.06		\\
	CCCP-HE		&	R	&	8	&	-0.13	&	0.15	&	0.38	&	0.13		\\
	\hline
	\end{tabular}
\end{table}

\begin{figure}
       \resizebox{\hsize}{!}{\includegraphics{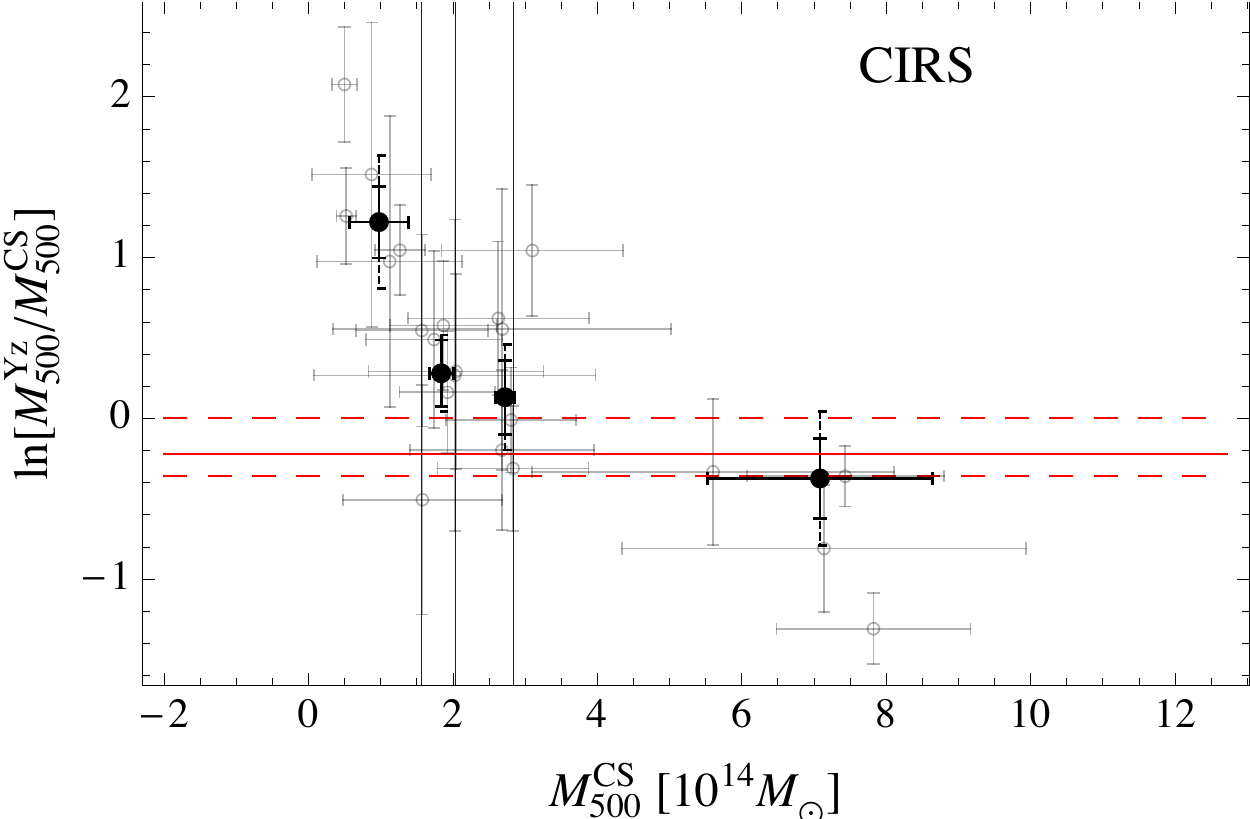}}
       \caption{Mass comparison, $\ln (M^{Y_z}_{500}/M^\mathrm{CS}_{500})$, for Planck clusters in the CIRS sample, as a function of $M^\mathrm{CS}_{500}$. Points and lines are as in Fig.~\ref{fig_bias_M500_WL}.}
	\label{fig_bias_M500_CS}
\end{figure}

\begin{figure*}
\begin{tabular}{cc}
\includegraphics[width=8.5cm]{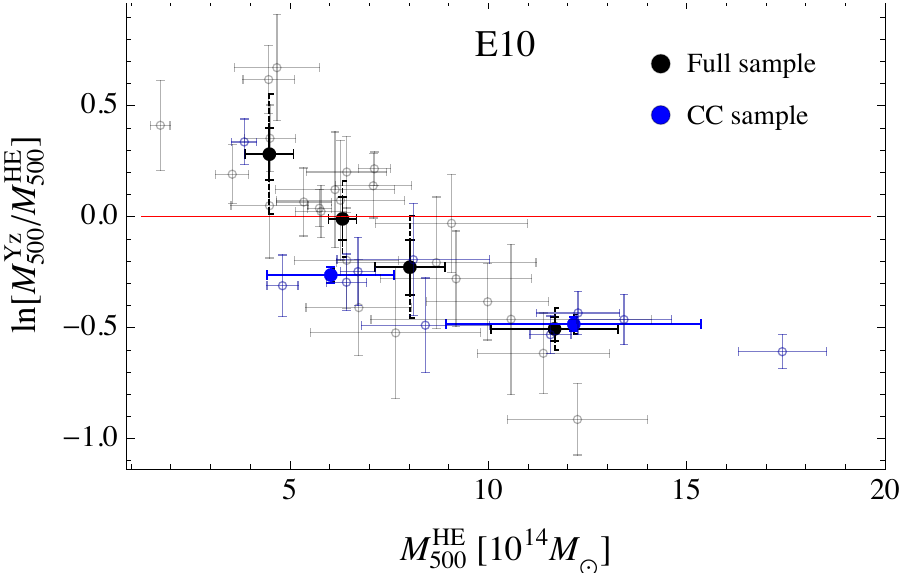} &\includegraphics[width=8.5cm]{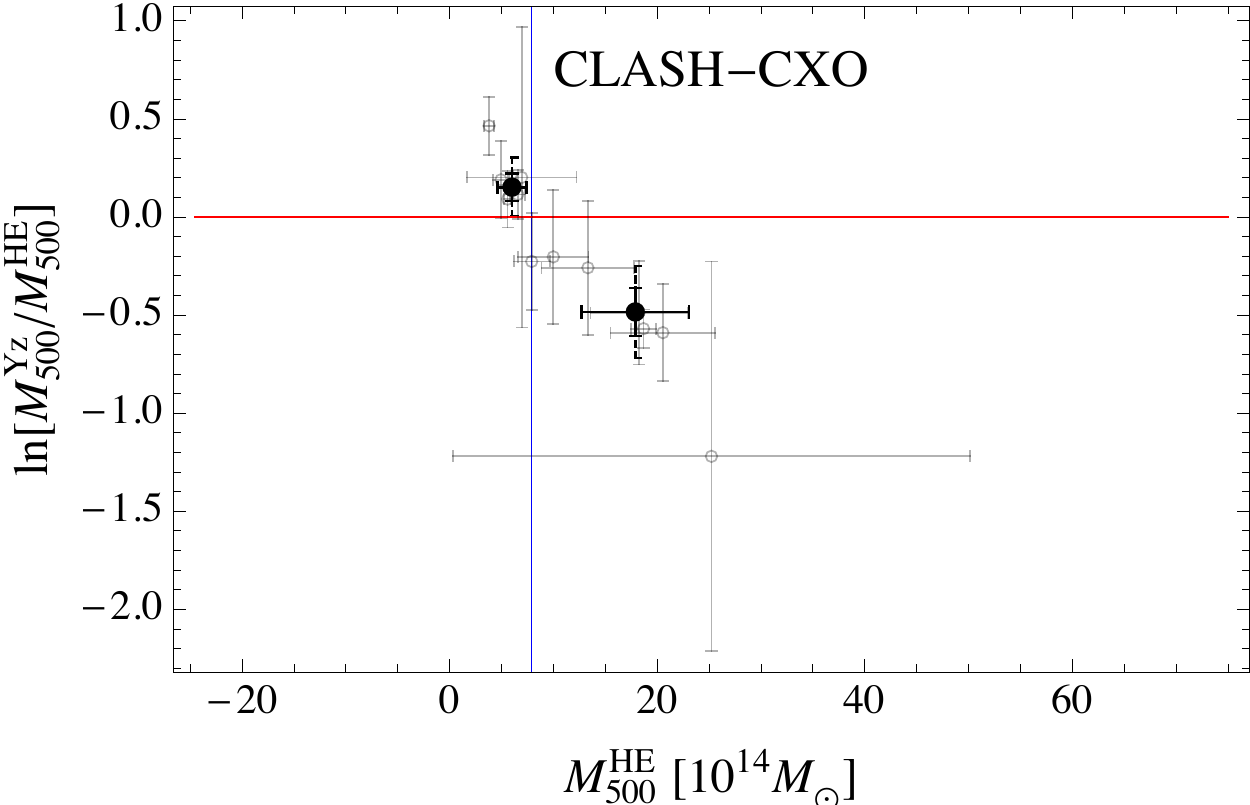}\\
\noalign{\smallskip}  
\includegraphics[width=8.5cm]{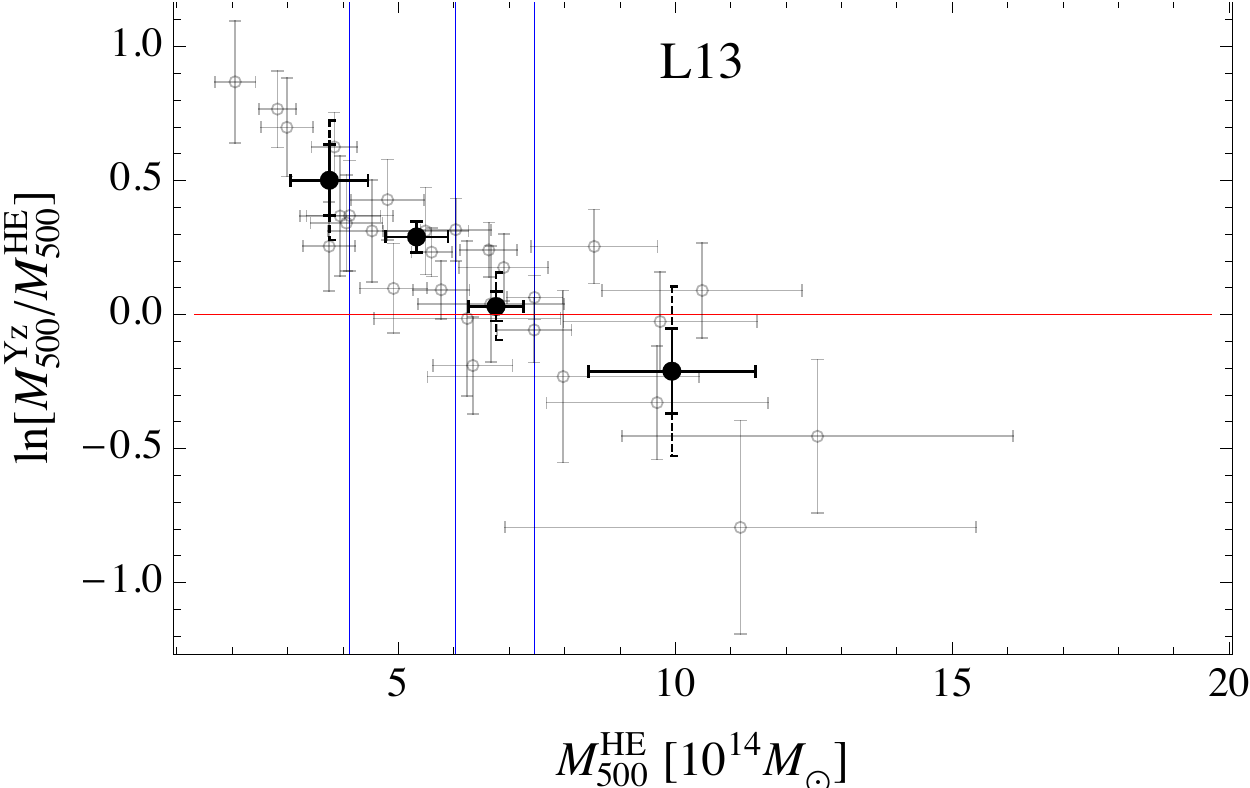} & \includegraphics[width=8.6cm]{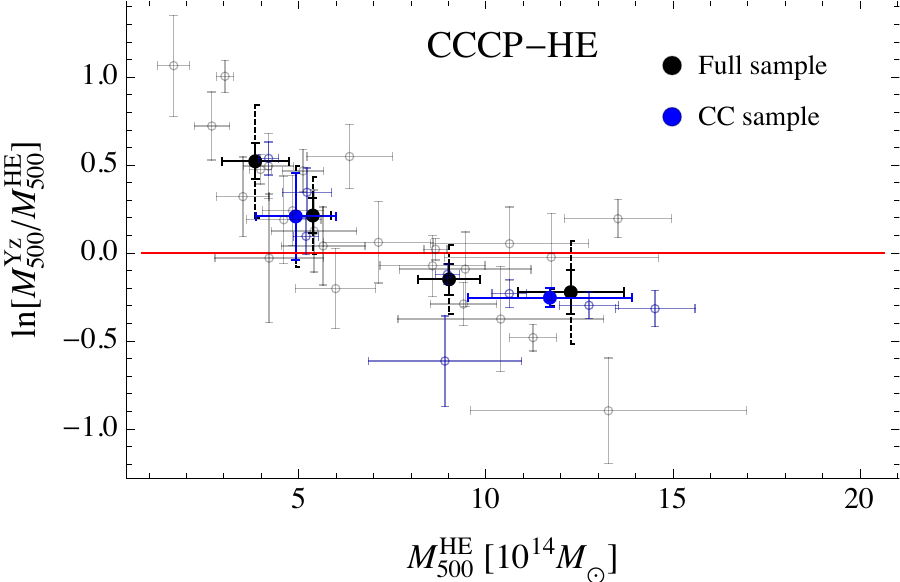}
\end{tabular}
\caption{
Mass comparison, $\ln (M^{Y_z}_{500}/M^\mathrm{HE}_{500})$, for Planck clusters with HE masses, as a function of $M^\mathrm{HE}_{500}$. Black and grey points, and blue lines are as in Fig.~\ref{fig_bias_M500_WL}. The red line plots a null bias. Blue points are the analog of the black ones but for the relaxed clusters. The top left, top right, bottom left, and bottom right panels show the results for the E10, the CLASH-CXO,  the L13, and the CCCP-HE samples, respectively.}
\label{fig_bias_M500_HE}
\end{figure*}

The inferred bias of the Planck masses $M^{Y_z}$ is relative to the adopted mass calibration sample they are compared against. We computed the bias for different calibration samples following the methodology detailed in \citetalias{se+et14_comalit_I}. We considered the (natural) logarithm of the unweighted mass ratios. Here and in the following, central estimates and scatters were calculated as bi-weight estimators. Uncertainties were estimated with bootstrap resampling with replacement. 

We performed the analysis for the WL samples  (see Fig.~\ref{fig_bias_M500_WL} and Table~\ref{tab_bias_M500_WL}), for the CS sample (see Fig.~\ref{fig_bias_M500_CS} and Table~\ref{tab_bias_M500_CS}), and for the X-ray samples (see Fig.~\ref{fig_bias_M500_HE} and Table~\ref{tab_bias_M500_HE}). Whatever the calibration sample, we found that the mass ratio is a decreasing function of the mass proxy, i.e., the bias $b^{Y_z}(\equiv 1- M^{Y_z}/M^\mathrm{Pr})$ steadily increases towards large masses. This feature is further highlighted when we bin clusters according to the mass. Estimates and trends do not change significantly after restricting the analysis to the cosmological sub-samples. 

There are several plausible sources for this trend. Firstly, as discussed in \citetalias{se+et14_comalit_I}, the WL, HE, and CS masses are scattered proxies of the true mass. 
Since the intrinsic scatters affecting $M^\mathrm{WL}$ and $M^{Y_z}$ are poorly correlated, for small (large) values of $M^\mathrm{WL}$ the estimate of $M^{Y_z}/M^\mathrm{WL}$ is biased high (low). The same considerations apply to the CS mass.

The HE mass and $M^{Y_z}$ can be correlated to some degree, being both connected to the pressure profile. However, departures from equilibrium, non-thermal contributions to the pressure, gas clumpiness, the presence of cool cores, and the efficiency of feedback processes affect HE masses and the SZ proxy in different ways. These two proxies can then be considered uncorrelated to first approximation.

Secondly, the mass dependent effect may suggest that the scaling relation used to infer $M^{Y_z}$ differs from the one characterising the calibration sample. As discussed in \citet{lin+al14}, a wrongly estimated slope $\beta$ could induce a mass dependent effect. We will see in Sec.~\ref{sec_scal_rela} that this is actually the case.

Thirdly, the Planck clusters are selected in SZ flux and suffer from Malmquist bias. Clusters with strong SZ emission are over-represented, mostly at low masses. The cluster mass obtained from the SZ flux through a scaling relation is then biased high near the flux threshold if no correction for Malmquist bias is applied to the detected flux or if this bias is not accounted for. Even if the scaling relation used to compute $M^{Y_z}$ was derived after correcting each flux for this bias, the measured values of $Y_{500}$ used to infer the cluster mass through the scaling relation should be corrected too.

Finally, the calibration masses may be systematically biased. We strongly disfavour this last hypothesis since the effect is common for all samples, whose masses were obtained with independent techniques.

The level of bias can not be ascertained with confidence due to the hidden systematics affecting mass determinations. As discussed in \citetalias{se+et14_comalit_I}, WL mass calibrations are still debated and differences in reported $M_{500}$ from independent analyses can be as large as 40 per cent, well beyond the level of known systematic errors plaguing WL analyses. As a consequence the estimation of the bias ranges from $\sim $ 0 (CCCP-WL) to $\gs$ 40 per cent (WTG and CLASH-WL). The bias for the WTG sample is in agreement with the result of \citet{lin+al14}.

The mass dependent effect shows up in the intrinsic scatter too (see Tables \ref{tab_bias_M500_WL}, \ref{tab_bias_M500_CS}, and  \ref{tab_bias_M500_HE}). The dispersion of mass ratios over too large mass intervals is artificially inflated if the assumed slope in the scaling relation for $M^{Y_z}$ is wrong. This over-estimates the intrinsic scatter for the full mass range. 

Let us consider the WL samples. Being the scatters uncorrelated, the total measured dispersion should be given by the contributions from the WL measurement errors ($\sim 15$ per cent), the intrinsic scatter in the WL mass, $\sim$ 10-15 per cent \citep{ras+al12,se+et14_comalit_I}, and the intrinsic scatter in the $M_{500}$-$Y_{500}$ relation, $\sim$ 15-20 per cent \citep{planck_2013_XX}. Adding in quadrature we expect the total dispersion to be of the order of $\sim$ 20-30 per cent, slightly smaller than the measured dispersion ($\sim$ 30-40 per cent).

If we confine the analysis to larger masses, the spread is smaller and the scatter is smaller too, see Fig.~\ref{fig_bias_M500_WL}. 

The mass bias found for clusters with caustic masses should behave as for WL clusters (see Fig.~\ref{fig_bias_M500_CS} and Table~\ref{tab_bias_M500_CS}). Notwithstanding the larger uncertainties, we retrieved the same features.

Complementary results can be obtained from the comparison of Planck masses to X-ray catalogs. The Planck $M^{Y_z}$ masses should reproduce the HE masses for relaxed clusters. Differently from the WL case, $M^{Y_z}$ is a tracker of the proxy $M^\mathrm{HE}$ and the bias $b^{Y_z}$ should be null, $M^{Y_z}- M^\mathrm{HE}\sim0$. 

Notwithstanding these differences with respect to the WL case, we found that the mass ratio $M^{Y_z}/M^\mathrm{HE}$ is a decreasing function of the mass proxy $M^\mathrm{HE}$ (see Fig.~\ref{fig_bias_M500_HE}). This strengthens the hypothesis that there is a mass dependent effect in the calibration of $M^{Y_z}$. 

An alternative hypothesis is that the different dynamical state of the considered X-ray clusters may induce some strong mass-dependent bias in the estimate of the HE mass. We considered subsamples of relaxed clusters for the E10 and the CCCP-HE catalogs (see Fig.~\ref{fig_bias_M500_HE} and Table~\ref{tab_bias_M500_HE}). We found that the mass-dependence of the bias is still in place whereas the level of bias is increased. However, these subsamples of relaxed clusters are small and we caution against over-interpretation.

\section{Planck calibration sample}
\label{sec_cali_samp}

\begin{figure*}
\begin{tabular}{cc}
\includegraphics[width=8.5cm]{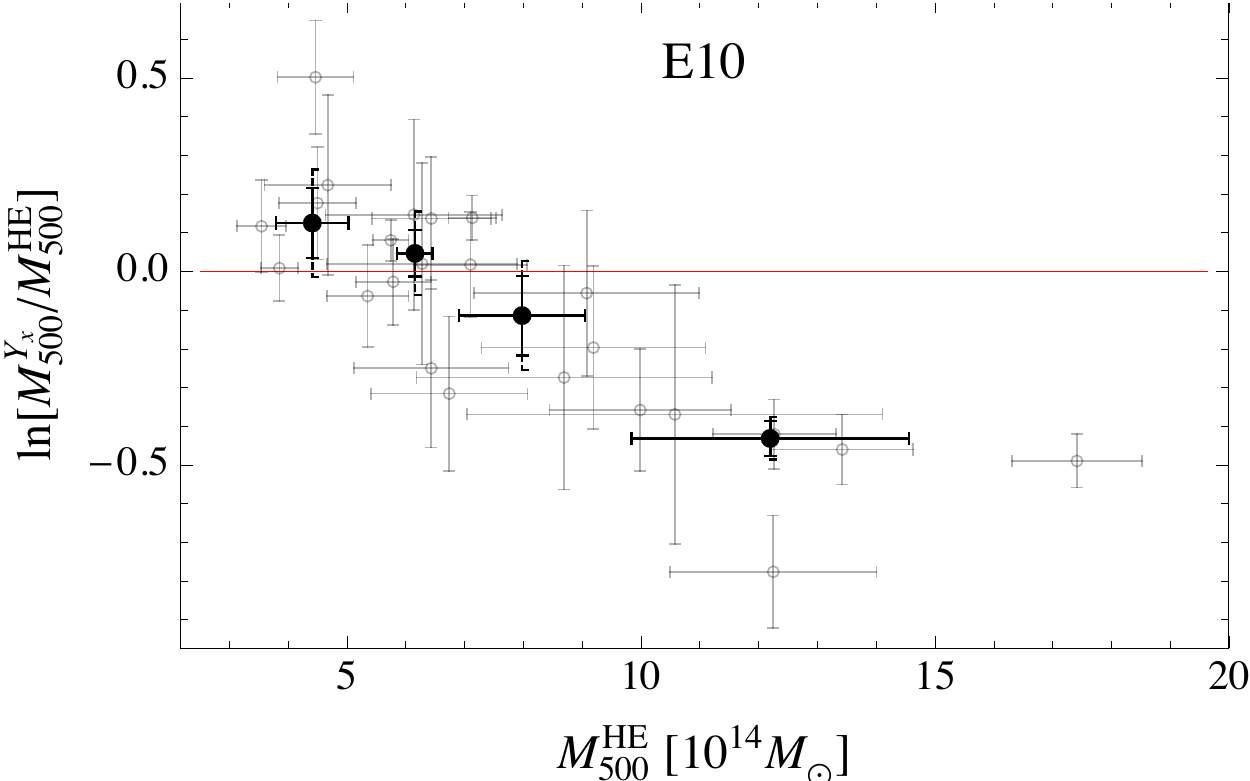} &\includegraphics[width=8.5cm]{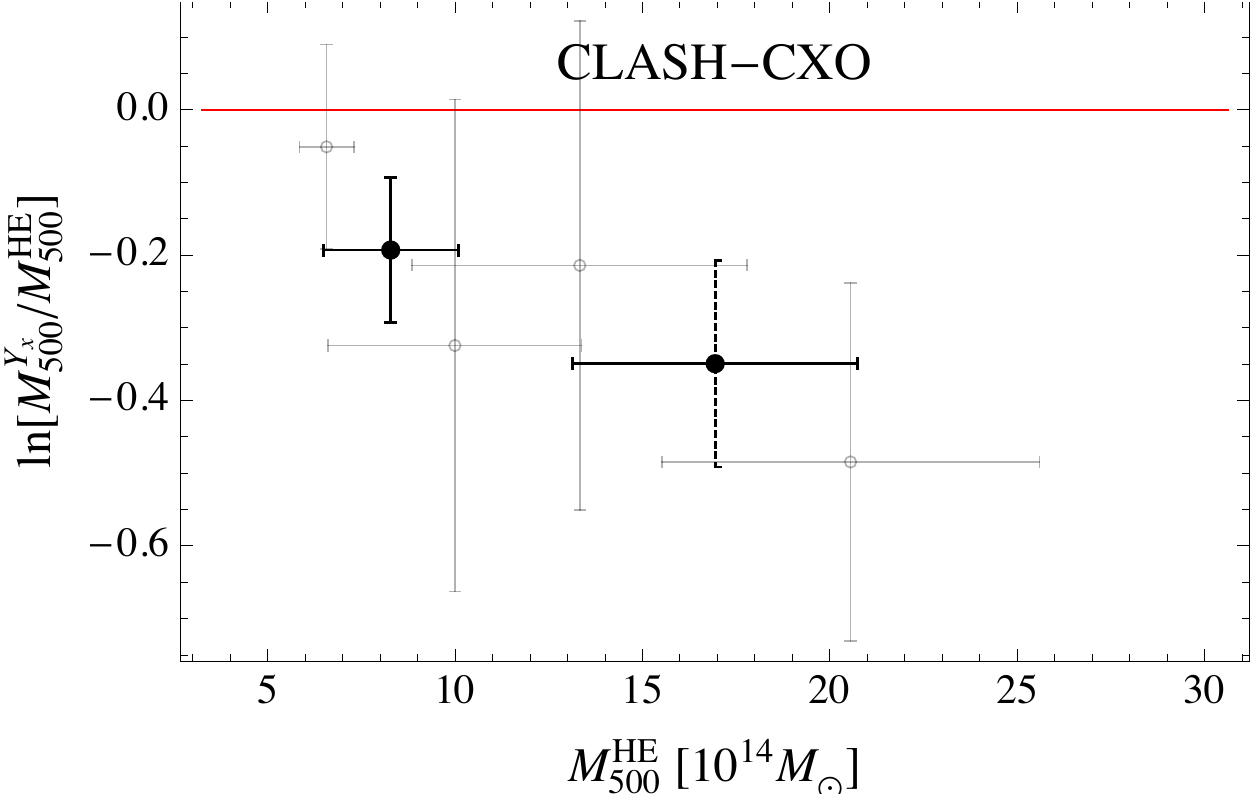}\\
\noalign{\smallskip}  
\includegraphics[width=8.5cm]{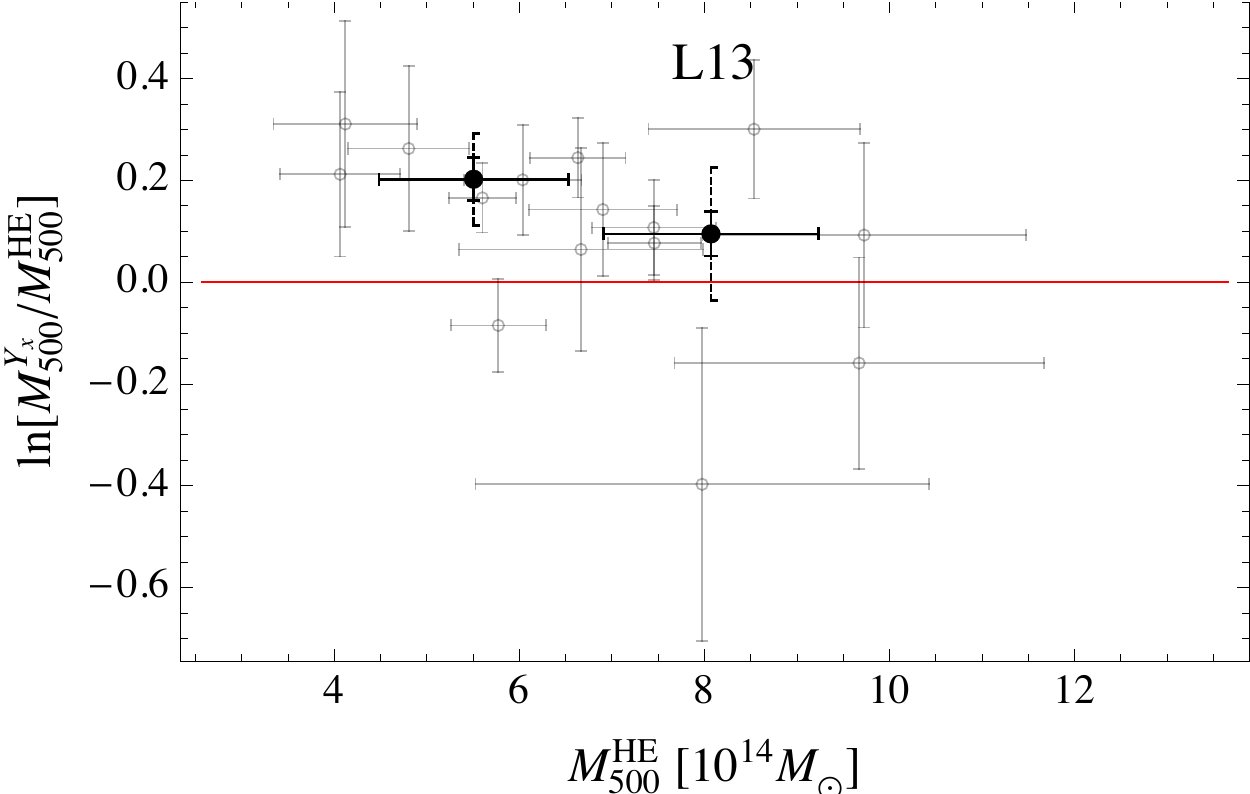} & \includegraphics[width=8.6cm]{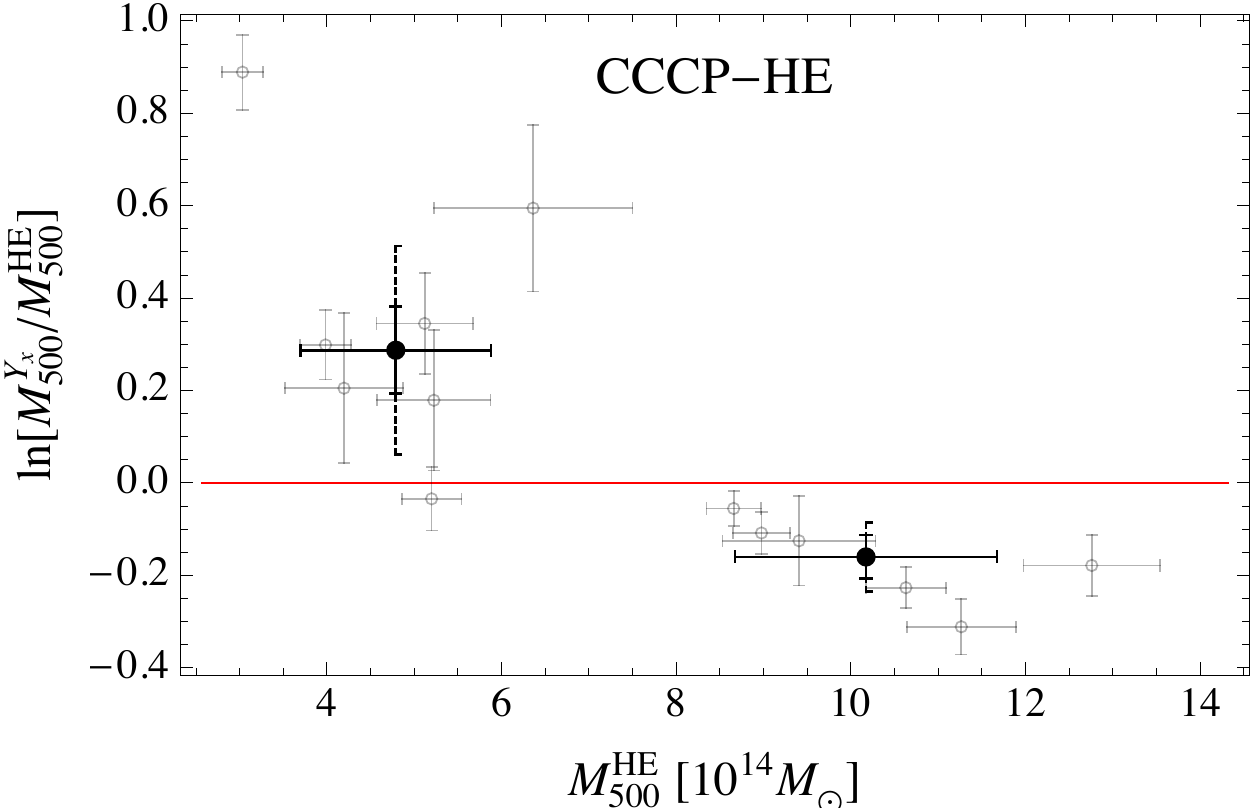}
\end{tabular}
\caption{
Mass comparison, $\ln (M^{Y_\mathrm{X}}_{500}/M^\mathrm{HE}_{500})$, for Planck clusters in the calibration sample with estimated HE masses, as a function of $M^\mathrm{HE}_{500}$. Points and lines are as in Fig.~\ref{fig_bias_M500_HE}. The red line plots a null bias. The top left, top right, bottom left, and bottom right panels show the results for the E10, the CLASH-CXO,  the L13, and the CCCP-HE samples, respectively.}
\label{fig_bias_MYX_HE}
\end{figure*}

\begin{table}
\caption{Mass comparison, $\ln M_{500}^{Y_\mathrm{X}} -\ln M_{500}^\mathrm{HE}$, for Planck clusters in the calibration samples. Col.~1: mass catalog used for the calibration. Col.~2: subsample (`C' refers to the Planck clusters in the cosmological subsample; `R' refers to the relaxed clusters ). Col.~3: number of clusters, $N_\mathrm{Cl}$. Cols.~4 and 5: typical $\ln M_{500}^{Y_\mathrm{X}}/M_{500}^\mathrm{HE}$ and scatter. Central estimates and scatters are computed as bi-weighted estimators.}
\label{tab_bias_MYX_HE}
\centering
\begin{tabular}[c]{l  c r r@{$\,\pm\,$}l  r@{$\,\pm\,$}l}
	\hline
	Calibration		&	Sample	&	$N_\mathrm{Cl}$	&	\multicolumn{2}{c}{$\mu$}	&	\multicolumn{2}{c}{$\sigma$}  \\
	\hline
	E10			&	C	&	24	&	-0.08	&	0.07	&	0.29	&	0.05		\\
	E10			&	R	&	4	&	\multicolumn{2}{c}{$\sim-0.46$}	&	\multicolumn{2}{c}{0.04}		\\
	\hline
	CLASH-CXO	&	C	&	4	&	-0.27	&	0.10	&	0.17	&	0.07		\\
	\hline
	L13			&	C	&	15	&	0.15		&	0.05	&	0.15	&	0.05		\\
	\hline
	CCCP-HE		&	C	&	13	&	0.04		&	0.13	&	0.35	&	0.09		\\
	CCCP-HE		&	R	&	5	&	-0.11	&	0.08	&	0.14	&	0.07		\\	
	\hline
	\end{tabular}
\end{table}

\begin{figure}
\resizebox{\hsize}{!} {\includegraphics{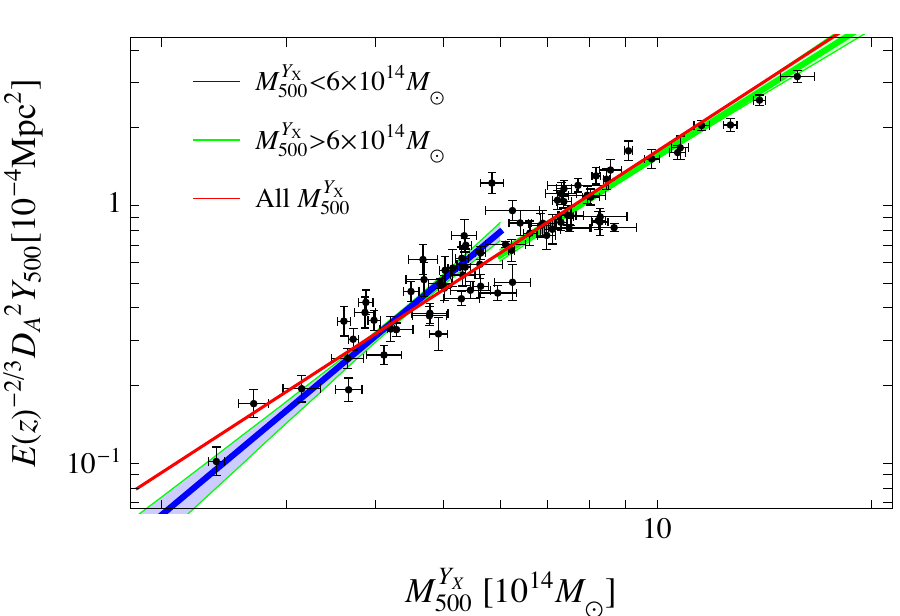}}
\caption{$M^{Y_\mathrm{X}}_{500}$ versus $Y_{500}$ for the Planck calibration sample. SZ fluxes are corrected for Malmquist bias. Regressions in different mass regimes, as obtained with a BCES-orthogonal method, are plotted. The blue (green) line considers the clusters with $M^{Y_\mathrm{X}}_{500}< (>)6\times10^{14}M_\odot$. Shaded regions are the 1-$\sigma$ confidence regions. The red line plots the fit for the full mass range.}
\label{fig_MYX_Y500_Planck}
\end{figure}

\begin{figure}
\resizebox{\hsize}{!} {\includegraphics{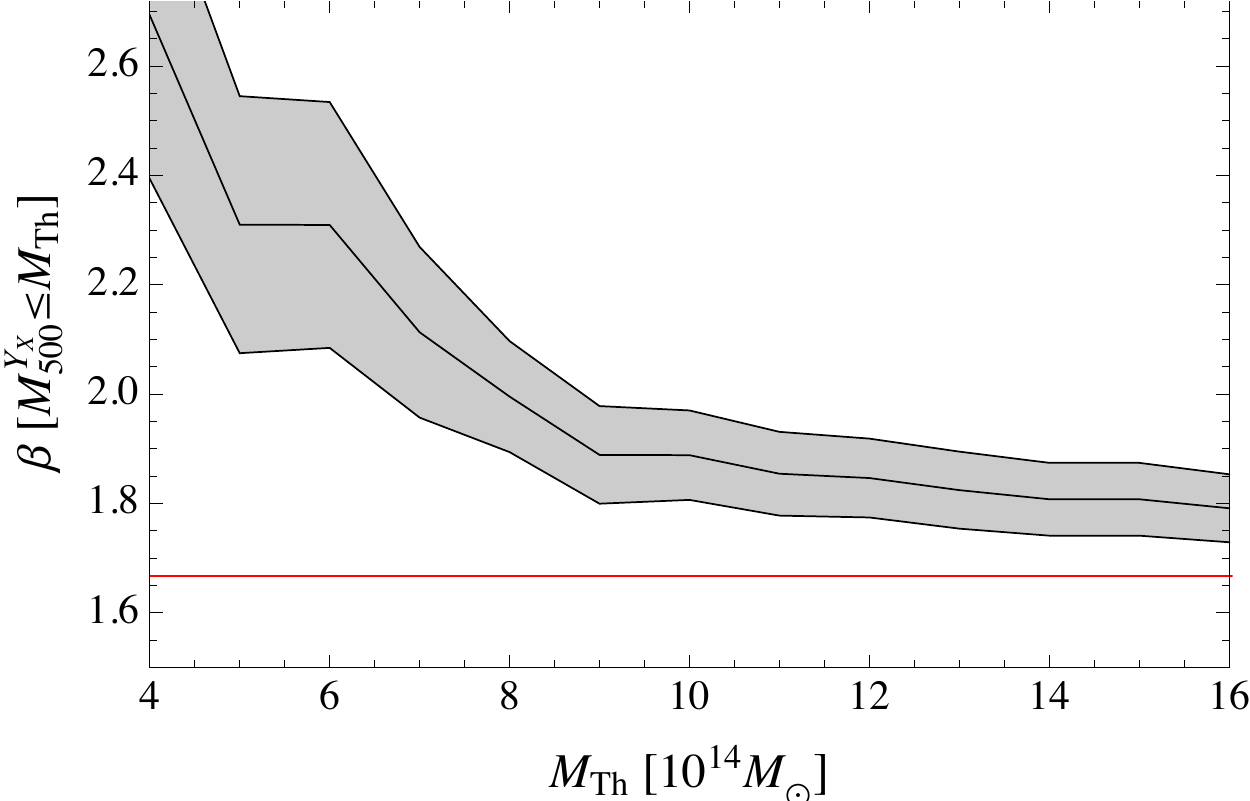}}
\caption{Slope $\beta$ of the scaling relation $M^{Y_\mathrm{X}}_{500}$-$Y_{500}$ as a function of the maximum mass $M_\mathrm{Th}$ of the considered clusters. We considered the 71 clusters in the Planck calibration sample and performed the linear regression with the BCES-orthogonal method after excising clusters whose mass proxy was larger than the threshold, $M^{Y_\mathrm{X}}_{500}>M_\mathrm{Th}$. The red line denotes the self-similar slope $\beta=5/3$.}
\label{fig_beta_BCES_vs_MThreshold_MYX}
\end{figure}

In this section, we show that the mass-dependent bias of the Planck masses found in Sec.~\ref{sec_mass_bias} is partially due to the biased estimates of $M^{Y_\mathrm{X}}$ in the Planck calibration sample. The determination of the Planck scaling relation $Y_{500}$-$M_{500}$ was obtained through a multiple step procedure. The two main pieces are the local $Y_\mathrm{X}$-$M^\mathrm{HE}_{500}$ and the $Y_{500}$-$Y_\mathrm{X}$ scaling relations. Consequently, the mass calibration based on the $Y_\mathrm{X}$ proxy, $M^{Y_\mathrm{X}}$, is at the core of the procedure. 

By construction the proxy $Y_\mathrm{X}$ mimics the SZ flux. The scatter between $Y_\mathrm{X}$ and $Y_{500}$ is due to cluster-to-cluster variations in the pressure and density profile \citep{arn+al10}, which are not so large. In fact, the slope of the $Y_{500}$-$M_{500}$ relation determined by the Planck team equals that of the $Y_\mathrm{X}$-$M_{500}$ relation to very good approximation.

Any bias in the estimate of $M^{Y_z}$ is then likely connected to systematics in the calibration of $M^{Y_\mathrm{X}}$. We repeated the comparison of Sec.~\ref{sec_mass_bias} to look for biases in the estimate of $M^{Y_\mathrm{X}}$ of the calibration sample of 71 clusters used to infer the $Y_{500}$-$M^{Y_\mathrm{X}}$ scaling relation.\footnote{We used data listed in the `MY\_4\_scaling.fits' catalog, which is publicly available at \url{http://szcluster-db.ias.u-psud.fr}.  The catalog reports the masses $M^{Y_\mathrm{X}}$, the SZ Compton parameters, and the cluster-by-cluster Malmquist bias corrections used in \citet{planck_2013_XX}.}

As $M^{Y_\mathrm{X}}$ was calibrated to reproduce the HE mass, we limited the analysis to the X-ray calibration samples. Results are summarised in Fig.~\ref{fig_bias_MYX_HE} and Table~\ref{fig_bias_MYX_HE}. We retrieved the same trends found for $M^{Y_z}$ (see Sec.~\ref{sec_mass_bias}). Differences between $M^{Y_\mathrm{X}}$ and $M^\mathrm{HE}$ are mass dependent. The mass ratio $M^{Y_\mathrm{X}}/M^\mathrm{HE}$ decreases with increasing masses. 

As discussed above, the assessment of the level of bias in the normalisation is hampered by the systematic differences in the calibration samples (see Table~\ref{fig_bias_MYX_HE}). The $M^{Y_\mathrm{X}}$ masses from Planck are biased low with respect to the E10 and CLASH-CXO samples, biased high with respect to the L13 sample, and consistent with the CCCP-HE sample. 

$M^{Y_\mathrm{X}}$ should be a good proxy whatever the equilibrium state of the cluster, whereas HE masses are biased low in disturbed systems. There are a few relaxed clusters in common among the various samples, which tentatively suggest that $M^{Y_\mathrm{X}}$ might be biased low for relaxed systems (see Table~\ref{fig_bias_MYX_HE}).

To further investigate whether the slope in the $Y_{500}$-$M_{500}$ is biased, we reanalysed the calibration sample. The Planck team adopted the BCES (Bivariate Correlated Errors and intrinsic Scatter) orthogonal regression method \citep{ak+be96}, which we use in this section to simplify comparison and stress the mass dependent effect. SZ fluxes were corrected for Malmquist bias as suggested in \citet{planck_2013_XX}. First, we checked that we re-obtained the same regression parameters when the whole sample of 71 clusters is considered ($\alpha=-0.186 \pm0.011$ and $\beta=1.79\pm0.06$ for $M_\mathrm{pivot}=6\times10^{14}M_\odot$). 

Quoted uncertainties in the Planck calibration sample are small and the data are well aligned. As a consequence, the estimated slope does not depend significantly on the regression method. The slope found with the Bayesian method detailed in \citetalias{se+et14_comalit_I} is $\beta \sim 1.7$. 

The value of the slope is strongly impacted by a few clusters with very large mass (see Fig.~\ref{fig_MYX_Y500_Planck}). The scaling relation estimated considering only the lower mass half of the clusters, $M^{Y_\mathrm{X}}_{500} <6 \times 10^{14}M_\odot$, has slope $\beta=2.29\pm0.26$; for the more massive half, $M^{Y_\mathrm{X}}_{500} >6 \times 10^{14}M_\odot$, the slope is $\beta=1.74\pm0.11$. 

The mass dependent effect for different mass thresholds is shown in Fig.~\ref{fig_beta_BCES_vs_MThreshold_MYX}. The flattening of the slope is driven by the most massive systems. The value of $\beta$ is approximately self-similar only due to a few very massive clusters with $M^{Y_\mathrm{X}}_{500} \gs 10 \times 10^{14}M_\odot$.

The Planck calibration sample consists of 71 detections from the Planck cosmological cluster sample, detected at $\mathrm{SNR}>7$, for which good quality XMM-Newton observations were available \citep{planck_2013_XX}. The heterogeneous nature of the sample may be at the origin of the very steep low mass slope.

\section{Scaling relation}
\label{sec_scal_rela}

In this section we describe how we performed the $Y_{500}$-$M_{500}$ regression. The value of $Y_{500}$ has to be consistently measured within the over-density radius corresponding to $M_{500}$. We first estimated the SZ fluxes and then we performed the regression based on the Bayesian technique detailed in \citetalias{se+et14_comalit_I}. The regression method was further developed by {\it i)} accounting for the covariance of the uncertainties, and {\it ii)} including a correction for the Malmquist bias.

\subsection{SZ signal}
\label{sec_SZ_sig}

To study the scaling relation between mass and SZ flux, we re-computed the spherically integrated $Y_{500}$ of the PSZ1 clusters with external WL, CS, or HE mass determinations within the radius $r_{500}$. The prior on $r_{500}$ reduces the size-flux degeneracy. As probability distribution, we used
\beq
p_\mathrm{PSZ1}(Y_{500}, r_{500}) \propto{\cal L}_\mathrm{PSZ1}(Y_{500}, r_{500})P(r_{500}),
\eeq
where ${\cal L}_\mathrm{PSZ1}(Y_{500}, r_{500})$ is the likelihood of $Y_{500}$ and $r_{500}$ as obtained by the Planck team with the Matched Multi-Filter method Three.\footnote{We retrieved the likelihood functions from the catalog `COM\_PCCS\_SZ-MMF3\_R1.12.fits', which is publicly available from the Planck Legacy Archive. The Planck team actually measured $Y_{5r_{500}}$ and the scaling radius $\theta_\mathrm{s}$ of the assumed pressure profile \citep{arn+al10}. Since they employed a profile with fixed concentration, the corresponding values of $Y_{500}$ and $r_{500}$ are proportional to $Y_{5r_{500}}$ and $\theta_\mathrm{s}$.}

$P(r_{500})$ is the Gaussian prior on $r_{500}$ as determined from the external information on $M_{500}$ (either from WL, X-ray, or caustic analyses). We computed the SZ signal as
\beq
\label{eq_Y500}
Y_{500} = \int Y^{'}_{500}\, p_\mathrm{PSZ1}(Y^{'}_{500}, r_{500})d r_{500}.
\eeq 
The full uncertainty covariance matrix was derived in a similar way by computing the second order momenta of the probability distribution. The posterior marginalised distribution of $r_{500}$ simply follows the prior.

We tested the above procedure against the $Y_z$ reported in the validation catalogue. The main difference between the computation of $Y_z$ reported in the Planck catalog or the values $Y_{500}$ computed as in Eq.~(\ref{eq_Y500}) consists in the assumed prior. $Y_z$ assumed the Planck $Y_{500}$-$M_{500}$ scaling relation. We assumed a prior on $r_{500}$ as derived from the mass knowledge.

To compare our estimate of $Y_{500}$ to $Y_z$, we then assumed as $r_{500}$ the radius derived from the listed $M^{Y_z}$. For the 664 clusters with known redshift detected with the MMF3 method, we found a very good agreement, $Y_{500}/Y_{z} \sim 1.02 \pm 0.02$. Breaking the size-flux degeneracy through the scaling relation rather than assuming a prior on $r_{500}$ from an external determination is slightly less constraining and produces a slightly larger uncertainty on the estimated $Y_{500}$. For the 664 clusters, we found $\delta Y_{500}/\delta Y_{z} \sim 0.91 \pm 0.05$.

Our estimates of $Y_{500}$  were based on the original detections. Nevertheless, the error determined by not re-centring the signal around the identified X-ray/optical counterpart is negligible. Firstly, we compared the original $Y_z$ to $Y_\mathrm{500,PSX}$, the signal re-extracted at the X-ray position fixing the size to the X-ray size, as provided in the Planck external validation catalog \citep{planck_2013_XXIX}. For the 869 clusters with complete information, we found $Y_\mathrm{500,PSX}/Y_z \sim 1.00 \pm 0.01$. The typical difference, $\Delta Y = Y_\mathrm{PSX} -Y_z$, is $\sim$20 per cent of the quoted error on $Y_z$.

Secondly, we compared our determination of $Y_{500}$ to the re-extracted values obtained by the Planck collaboration for their calibration cosmological sample of 71 clusters. For consistency, as prior for $r_{500}$ we used the estimated $r^{Y_\mathrm{X}}_{500}$, i.e., the radius corresponding to $M^{Y_\mathrm{X}}$. We found $Y_{500}/Y_\mathrm{500,PSX} \sim 1.0 \pm 0.1$. The typical deviation, $\Delta Y= Y_\mathrm{500,PSX} -Y_{500}$, is $\sim -0.1$ times the uncertainty on $Y_{500}$. The estimates of the uncertainties agree too. we found $\delta Y_{500}/\delta Y_\mathrm{500,PSX} \sim 1.07 \pm 0.12$.

We can then conclude that our procedure to compute $Y_{500}$ is reliable and that re-extracting and re-centring the signal would not have significantly impacted our results.

\subsection{Regression method}

The regression was implemented through the Bayesian method detailed in \citetalias{se+et14_comalit_I}. We modelled the $Y_{500}$-$M_{500}$ relation with a single power law, see Eq.~(\ref{eq_scal_rela}).  The linear regression was performed in decimal logarithmic variables,
\begin{eqnarray}
\log \tilde{Y}_{500}	\pm \delta_{\log \tilde{Y}_{500}}	& \sim & \alpha_{Y|Z} + \beta_{Y|Z} \log \tilde{M}_{500} \pm \sigma_{Y|Z}  \label{eq_bug_1}\\
\log M^\mathrm{Pr}	\pm \delta_{\log M^\mathrm{Pr}}	& \sim &  \log M_{500} \pm \sigma_{X|Z}  \label{eq_bug_2}
\end{eqnarray}
where $\tilde{Y}_{500} \equiv (H(z)/H_0)^{-2/3} (D^2(z)Y_{500}/ 10^{-4}\mathrm{Mpc}^2)$ and $\tilde{M}_{500}  \equiv M_{500}/M_\mathrm{pivot}$. To easy the comparison, we kept the pivot mass fixed to $M_\mathrm{pivot}=6\times10^{14}M_\odot$ for all samples. According to the notation adopted in the CoMaLit series, see sec.~2.2 of \citetalias{se+et14_comalit_I}, we can identify the response variable $Y$ with $\log \tilde{Y}_{500}$, the scattered covariate variable $X$ with $\log (M_{500}^\mathrm{Pr}/M_\mathrm{pivot})$, and the hidden variable $Z$ with the true mass ($\log M_{500}/M_\mathrm{pivot}$). 

We assumed that the proxy mass is an unbiased but scattered tracer of the true mass, see Eq.~(\ref{eq_bug_2}). The statistical scheme summarised in Eqs.~(\ref{eq_bug_1}) and (\ref{eq_bug_2}) can only determine the relative bias between the mass proxies, i.e., between $Y$ and $X$, whereas the calibration of the absolute bias against the true mass needs additional information usually available only in simulations, see \citetalias{se+et14_comalit_I}.

The intrinsic scatters $\sigma_{Y|Z}$ and $ \sigma_{X|Z}$ refer to the dispersion of the SZ signal and of the mass proxy around the true mass, respectively. They are assumed to be uncorrelated and distributed according to log-normal functions as tested for optical \citep{sar+al13} or X-ray/SZ observables \citep{sta+al10,fab+al11}.

The intrinsic observable ($\log \tilde{Y}_{500}$ and $\log M^\mathrm{Pr}$) and the actually observed ($\log \tilde{Y}_{500}^\mathrm{obs}$ and $\log M^\mathrm{Pr,obs}$) fluxes and masses differ for the observational uncertainties ($\delta_{\log \tilde{Y}_{500}}$ and $\delta_{\log M^\mathrm{Pr}}$). Since the SZ flux is computed within $r_{500}$, the measurements of $Y_{500}$ and $M_{500}$ are correlated. We assume that the observed mass and SZ flux are randomly distributed following a joint bivariate distribution centred in the mass and flux we should measure with infinitely accurate and precise measurements and with uncertainty covariance matrix as computed in Sec.~\ref{sec_SZ_sig}. The diagonal element of the covariance matrix $\mathbfss{C}$ are $\delta_{\log \tilde{Y}_{500}}^2$ and $\delta_{\log \tilde{M}^\mathrm{Pr}}^2$. The off-diagonal elements are given by $\tilde{\rho}\ \delta_{\log \tilde{Y}_{500}} \delta_{\log \tilde{M}^\mathrm{Pr}}$, where $\tilde{\rho}$ is the correlation factor. This approach accounts for heteroscedastic and correlated measurement errors.

To account for selection effects, the distribution of the observed flux given a true latent value was truncated. The Malmquist bias, where bright objects near the flux limit are preferentially detected, was modelled by including a step function in the distribution of observed SZ flux. The threshold was SNR$>\mathrm{SNR_{Th}}=4.5$ (7) for the full (cosmological) sample. The probability of observed values can then summarised as
\begin{multline}
P(\log \tilde{Y}_{500,i}^\mathrm{obs},\log \tilde{M}^\mathrm{Pr, obs}_i) = {\cal N}_\mathrm{2D}(\log \tilde{Y}_{500,i}, \log \tilde{M}^\mathrm{Pr}_i; \mathbfss{C}_i) \\  \times
{\cal U}(\mathrm{SNR_{Th}}  \delta_{\log \tilde{Y}_{500}},\infty),
\end{multline}
where ${\cal N}_\mathrm{2D}$ and ${\cal U}$ are the bivariate Gaussian and the uniform distributions, respectively. 

The intrinsic distribution of the independent variable $Z$($=\log \tilde{M}_{500} $) was approximated with a Gaussian function of mean $\mu_Z$ and standard deviation $\sigma_Z$, 
\beq
Z_i 	\sim {\cal N} (\mu_Z, \sigma_Z).
\eeq
This model is in agreement with the observed mass distributions (see Fig.~\ref{fig_histo_M500_all}), and it is suitable for flux selected samples of massive clusters, see \citetalias{se+et14_comalit_IV}.  Alternatively, we approximated the intrinsic distribution of the independent variable using a mixture of Gaussian functions \citep{kel07},
\beq
Z_i 	\sim \sum_{k=1}^{n_\mathrm{mix}}\mathcal{\pi}_k {\cal N} (\mu_{Z,k}, \sigma_{Z,k}),
\eeq
where $n_\mathrm{mix}$ is the total number of Gaussian functions, and $\mathcal{\pi}_k$ is the probability of drawing a data point from the $k$th Gaussian function; $\sum_{k=1}^{n_\mathrm{mix}}\mathcal{\pi}_k =1$. We checked that the impact of modelling the intrinsic distribution with a mixture of Gaussian functions is minimal and we adopted the simpler hypothesis as reference case. The Gaussian function is flexible enough to mimic a nearly uniform distribution in case of very large variance.

We chose priors as less informative as possible, see \citetalias{se+et14_comalit_I}. We adopted uniform priors for the intercept $\alpha_{Y|Z}$, and the mean $\mu_Z$,
\beq
\alpha_{Y|Z},\  \mu_Z  \sim  {\cal U}(-1/\epsilon,1/\epsilon),
\eeq
where $\epsilon$ is a small number. In our calculation we took $\epsilon = 10^{-3}$. For the inverse of the variances, $\sigma_{Y|Z}^{-2}$, $\sigma_{X|Z}^{-2}$, and $\sigma_Z^{-2}$, we considered Gamma distributions \citep{an+hu12},
\beq
1/\sigma_{X|Z}^2,\ 1/\sigma_{Y|Z}^2,\ 1/\sigma_{Z}^2 \sim \Gamma(\epsilon,\epsilon),
\eeq
For the slope $\beta_{Y|Z}$, we assumed a Student's $t_1$ distribution, which is equivalent to a uniform prior on the direction angle $\arctan \beta_{Y|Z}$,
\beq
\beta_{Y|Z} \sim  t_1.
\eeq
When the intrinsic distribution of the independent variable was modelled with a Gaussian mixture, we modelled the prior distributions of the parameters $\mu_{Z,k}$ and $\sigma_{Z,k}$ as those of $\mu_{Z}$ and $\sigma_{Z}$, respectively. For the mixture probability coefficients, we adopted a Dirichelet distribution \citep{kel07},
\beq
\mathcal{\pi}_1, ..., \mathcal{\pi}_{n_\mathrm{mix}} \sim \mathrm{Dirichelet}(1, ..., 1),
\eeq
which is equivalent to a uniform prior on the $\mathcal{\pi}_k$'s under the constraint  $\sum_{k=1}^{n_\mathrm{mix}}\mathcal{\pi}_k =1$. The number of Gaussian functions in the mixture $n_\mathrm{mix}$ was fixed.

Some of the above priors differ from the usual hidden priors adopted in linear fitting procedure. The prior that the slope (instead of the direction angle) is uniformly distributed biases the estimate of $\beta_{Y|Z}$ high. At the same time, neglecting positive correlation between measured quantities can as well bias the slope high. On the other hand, assuming that the true masses are uniformly distributed biases the slope down if the mass distribution is fairly peaked, see \citetalias{se+et14_comalit_I}. 

For the numerical implementation of the above statistical scheme, we used the software JAGS (Just Another Gibbs sampler)\footnote{JAGS is publicly available at \url{http://mcmc-jags.sourceforge.net}.}. We verified that the alternative statistical approach described in \citet{kel07} gives results in full agreement with the Bayesian method detailed above when we neglect the intrinsic scatter in the mass proxy and if we uniform the treatment of the priors and of the Malmquist bias.

\subsection{Slopes}
\label{sec_slop}

A procedure of linear regression may answer to different questions. We refer to \citet{iso+al90}, \citet{ak+be96}, \citet{hog+al10}, \citet{an+hu12}, \citet{fe+ba12}, and references therein for different views with an astronomical-friendly language on the topic. We might be looking for the parameters that better describe the relation between two quantities $Y$ and $Z$ (the `symmetric' scaling relation),
\beq
\label{eq_slop_1}
Y +\beta_{Y\textrm{-}Z} Z = \alpha_{Y\textrm{-}Z},
\eeq
or we might try to predict the unknown value of quantity given the known value of a second one (the `conditional' or `predictive' scaling relation),
\beq
\label{eq_slop_2}
Y = \alpha_{Y|Z}+\beta_{Y|Z} Z .
\eeq
If the quantities aligned exactly, the two questions would have the same answer. In astronomy, relations among observable quantities are usually scattered and we have to face two distinct statistical problems. If we are studying the physical processes behind the formation and evolution of clusters we are interested in the parameters in Eq.~(\ref{eq_slop_1}). For example, if we want to study how mass and luminosity, or temperature, or flux are related and how they evolve we should employ regression tools such as the BCES-orthogonal estimator  \citep{ak+be96} or the method detailed in \citet{hog+al10}. 

If we want to use a easily measured quantity (such as X-ray or optical luminosity) to estimate a more elusive property (such as the mass), or if we want to compare measured properties (such as luminosity functions or other number counts) with theoretical prediction based on the halo mass function we are asking the second question and we are interested in the parameters in Eq.~(\ref{eq_slop_2}). The ordinary least squares estimator, the BCES estimator $\beta(X_2|X_1)$ \citep{ak+be96}, the LINMIX\_ERR routine \citep{kel07}, or other Bayesian methods \citep{an+hu12} can be employed in this regard.

In the statistical framework we are employing, wherein the scatter is normal and the distribution of the covariate variable $Z$ is Gaussian too, it is easy to determine the parameters of one scaling relation given the other one, see App.~\ref{app_slop}. When we will discuss the degree of self-similarity of the $Y_{500}$-$M_{500}$, we will consider the symmetric relation. When we will argue on the cosmological implications of the number counts of SZ clusters, we will consider the conditional relation.

\section{Results}
\label{sec_resu}

\begin{table*}
\caption{Parameters of the scaling relation $Y_{500}$-$M_{500}$ based on WL samples. Col. 1: sample used for mass calibration; col. 2: C or F denotes the cosmological or the full SZ sample, respectively; col. 3: number of clusters in the sample, $N_\mathrm{cl}$; col.~4: $Z$ is variable we calibrate against; cols. 5 and 6: intercept and slope of the conditional scaling relation; col 7: $\sigma_{Y|Z}$, intrinsic scatter of $Y_{500}$ with respect to the fitted scaling relation $Y_{500}$-$M_{500}$; col. 8: $\sigma_{X|Z}$, intrinsic scatter of the proxy mass with respect to the true mass $M_{500}$; col. 9:  Spearman's rank correlation coefficient $\rho_{\sigma}$ between $\sigma_{X|Z}$ and $\sigma_{Y|Z}$; cols. 10 and 11: intercept and slope of the symmetric scaling relation. Listed values of the scatters refer to the logarithm to base 10. Values in square brackets were fixed in the regression procedure. Quoted values are bi-weight estimators of the posterior probability distribution.}
\label{tab_M500_YSZ_WL}
\begin{tabular}[c]{l c r c r@{$\,\pm\,$}lr@{$\,\pm\,$}lr@{$\,\pm\,$}lr@{$\,\pm\,$}lrr@{$\,\pm\,$}lr@{$\,\pm\,$}l}
        \hline
        \noalign{\smallskip}
	Mass calibration	&	Sample	& $N_\mathrm{cl}$	&	$Z$	& \multicolumn{2}{c}{$\alpha_{Y|Z}$} & \multicolumn{2}{c}{$\beta_{Y|Z}$} & \multicolumn{2}{c}{$\sigma_{Y|Z}$} & \multicolumn{2}{c}{$\sigma_{X|Z}$} & $\rho_{\sigma}$& \multicolumn{2}{c}{$\alpha_{Y\textrm{-}Z}$} & \multicolumn{2}{c}{$\beta_{Y\textrm{-}Z}$} \\
        \noalign{\smallskip}
        \hline
	LC$^2$-{\it single}	&	F	&	115	&	$\log M_{500}$&				-0.28&	0.03	&	1.22	&	0.20	&	0.11	&	0.05	&	0.10	&	0.04			&$-0.78$	&-0.39&	0.03&	1.37&	0.15	\\
	LC$^2$-{\it single}	&	F	&	115	&	$\log M_{500}^\mathrm{WL}$&	-0.26&	0.02	&	1.00	&	0.10	&	0.16	&	0.02	&	\multicolumn{2}{c}{[0]}	&-- 		&-0.29&	0.03&	1.25&	0.13\\
	LC$^2$-{\it single}	&	C	&	64	&	$\log M_{500}$&				-0.26&	0.05	&	1.27	&	0.21	&	0.10	&	0.04	&	0.09	&	0.04			&$-0.68$	&-0.28&	0.05&	1.42&	0.19\\
	LC$^2$-{\it single}	&	C	&	64	&	$\log M_{500}^\mathrm{WL}$&	-0.23&	0.04	&	1.06	&	0.13	&	0.15	&	0.02	&	\multicolumn{2}{c}{[0]}	&-- 		&-0.27&	0.04&	1.31&	0.16\\
	\hline
	WTG			&	F	&	34	&	$\log M_{500}$&				-0.28&	0.09	&	1.14	&	0.28	&	0.07	&	0.03	&	0.06	&	0.03			&$-0.31$	&-0.32&	0.10&	1.29&	0.30\\
	WTG			&	F	&	34	&	$\log M_{500}^\mathrm{WL}$&	-0.24&	0.07	&	1.00	&	0.21	&	0.09	&	0.02	&	\multicolumn{2}{c}{[0]}	&--		&-0.30&	0.09&	1.20&	0.26\\
	WTG			&	C	&	22	&	$\log M_{500}$&				-0.24&	0.14	&	1.09	&	0.39	&	0.07	&	0.03	&	0.07	&	0.03			&$-0.28$	&-0.32&	0.17&	1.32&	0.47\\
	WTG			&	C	&	22	&	$\log M_{500}^\mathrm{WL}$&	-0.17&	0.10	&	0.87	&	0.27	&	0.10	&	0.03	&	\multicolumn{2}{c}{[0]}	&--		&-0.27&	0.13&	1.16&	0.37\\
	\hline
	CLASH-WL		&	F	&	11	&	$\log M_{500}$&				-0.31&	0.31	&	1.25	&	0.91	&	0.12	&	0.07	&	0.07	&	0.04			&$0.08$	&-0.64&	0.52&	2.24&	1.49\\
	CLASH-WL		&	F	&	11	&	$\log M_{500}^\mathrm{WL}$&	-0.31&	0.25	&	1.22	&	0.69	&	0.12	&	0.06	&	\multicolumn{2}{c}{[0]}	&--		&-0.57&	0.37&	2.00&	1.03\\
	CLASH-WL		&	C	&	6	&	$\log M_{500}$&				 0.08&	0.41	&	0.37	&	1.02	&	0.11	&	0.06	&	0.10	&	0.06			&$-0.05$	&-0.21&	2.12&	1.09&	5.40\\
	CLASH-WL		&	C	&	6	&	$\log M_{500}^\mathrm{WL}$&	 0.06&	0.29	&	0.42	&	0.72	&	0.12	&	0.05	&	\multicolumn{2}{c}{[0]}	&--		&-0.29&	1.07&	1.29&	2.69\\
	\hline
	CCCP-WL			&	F	&	35	&	$\log M_{500}$&				-0.22&	0.06	&	1.56	&	0.40	&	0.10	&	0.05	&	0.07	&	0.03			&$-0.49$	&-0.25&	0.07&	1.88&	0.42\\
	CCCP-WL			&	F	&	35	&	$\log M_{500}^\mathrm{WL}$&	-0.19&	0.05	&	1.23	&	0.25	&	0.15	&	0.03	&	\multicolumn{2}{c}{[0]}	&--		&-0.24&	0.06&	1.74&	0.37\\
	CCCP-WL			&	C	&	19	&	$\log M_{500}$&				-0.11&	0.08	&	1.20	&	0.48	&	0.08	&	0.04	&	0.07	&	0.03			&$-0.22$	&-0.17&	0.10&	1.57&	0.61\\
	CCCP-WL			&	C	&	19	&	$\log M_{500}^\mathrm{WL}$&	-0.08&	0.06	&	0.93	&	0.32	&	0.10	&	0.03	&	\multicolumn{2}{c}{[0]}	&--		&-0.14&	0.09&	1.38&	0.50\\
	\hline
	\end{tabular}
\end{table*}

\begin{table*}
\caption{Parameters of the scaling relation $Y_{500}$-$M_{500}$ based on the CIRS sample. The proxy mass is $M_{500}^\mathrm{CS}$. Columns are as in Table~\ref{tab_M500_YSZ_WL}.}
\label{tab_M500_YSZ_CS}
\begin{tabular}[c]{l c r c r@{$\,\pm\,$}lr@{$\,\pm\,$}lr@{$\,\pm\,$}lr@{$\,\pm\,$}lrr@{$\,\pm\,$}lr@{$\,\pm\,$}l}
        \hline
        \noalign{\smallskip}
	Mass calibration	&	Sample	& $N_\mathrm{cl}$	&	$Z$	& \multicolumn{2}{c}{$\alpha_{Y|Z}$} & \multicolumn{2}{c}{$\beta_{Y|Z}$} & \multicolumn{2}{c}{$\sigma_{Y|Z}$} & \multicolumn{2}{c}{$\sigma_{X|Z}$}& $\rho_{\sigma}$& \multicolumn{2}{c}{$\alpha_{Y\textrm{-}Z}$} & \multicolumn{2}{c}{$\beta_{Y\textrm{-}Z}$} \\
        \noalign{\smallskip}
        \hline
	CIRS			&	F	&	22	&	$\log M_{500}$&			-0.51&	0.45	&	0.71	&	1.03	&	0.40	&	0.13	&	0.24	&	0.10			&$-0.08$	&0.67&	2.21&	3.55&	5.36\\
	CIRS			&	F	&	22	&	$\log M_{500}^\mathrm{CS}$&	-0.60&	0.18	&	0.49	&	0.36	&	0.42	&	0.09	&	\multicolumn{2}{c}{[0]}	&--		&0.14&	0.76&	2.28&	1.81\\
	CIRS			&	C	&	10	&	$\log M_{500}$&			-0.31&	0.29	&	0.57	&	1.04	&	0.20	&	0.09	&	0.20	&	0.11			&$-0.15$	&0.02&	1.18&	1.88&	4.35\\
	CIRS			&	C	&	10	&	$\log M_{500}^\mathrm{CS}$&	-0.36&	0.14	&	0.35	&	0.41	&	0.23	&	0.06	&	\multicolumn{2}{c}{[0]}	&--		&0.20&	0.38&	1.00&	1.41\\
	\hline
	\end{tabular}
\end{table*}

\begin{table*}
\caption{Parameters of the scaling relation $M_{500}-Y_{500}$ based on X-ray samples. The proxy mass is the HE mass. Columns are as in Table~\ref{tab_M500_YSZ_WL}.}
\label{tab_M500_YSZ_HE}
\begin{tabular}[c]{l c r c r@{$\,\pm\,$}lr@{$\,\pm\,$}lr@{$\,\pm\,$}lr@{$\,\pm\,$}lrr@{$\,\pm\,$}lr@{$\,\pm\,$}l}
        \hline
        \noalign{\smallskip}
	Mass calibration	&	Sample	& $N_\mathrm{cl}$	&	$Z$	& \multicolumn{2}{c}{$\alpha_{Y|Z}$} & \multicolumn{2}{c}{$\beta_{Y|Z}$} & \multicolumn{2}{c}{$\sigma_{Y|Z}$} & \multicolumn{2}{c}{$\sigma_{X|Z}$} & $\rho_{\sigma}$&  \multicolumn{2}{c}{$\alpha_{Y\textrm{-}Z}$} & \multicolumn{2}{c}{$\beta_{Y\textrm{-}Z}$} \\
        \noalign{\smallskip}
        \hline
	E10				&	F	&	34	&	$\log M_{500}$				&	-0.22&	0.04	&	1.15	&	0.34	&	0.11	&	0.05	&	0.10	&	0.04			&$-0.61$	&-0.24&	0.05&	1.43&	0.33\\
	E10				&	F	&	34	&	$\log M_{500}^\mathrm{HE}$	&	-0.20&	0.03	&	0.84	&	0.17	&	0.15	&	0.03	&	\multicolumn{2}{c}{[0]}	&--		&-0.23&	0.04&	1.22&	0.26\\
	E10				&	C	&	27	&	$\log M_{500}$				&	-0.16&	0.04	&	0.93	&	0.37	&	0.09	&	0.04	&	0.10	&	0.04			&$-0.50$	&-0.19&	0.05&	1.21&	0.40	\\
	E10				&	C	&	27	&	$\log M_{500}^\mathrm{HE}$	&	-0.14&	0.03	&	0.63	&	0.17	&	0.12	&	0.02	&	\multicolumn{2}{c}{[0]}	&--		&-0.16&	0.04&	0.93&	0.24\\
	\hline
	CLASH-CXO		&	F	&	12	&	$\log M_{500}$				&	-0.11&	0.05	&	0.88	&	0.17	&	0.05	&	0.03	&	0.06	&	0.03			&0.11	&-0.12&	0.06&	0.92&	0.18\\
	CLASH-CXO		&	F	&	12	&	$\log M_{500}^\mathrm{HE}$	&	-0.10&	0.04	&	0.85	&	0.13	&	0.05	&	0.02	&	\multicolumn{2}{c}{[0]}	&--		&-0.11&	0.04&	0.88&	0.13\\
	CLASH-CXO		&	C	&	7	&	$\log M_{500}$				&	-0.06&	0.07	&	0.80	&	0.18	&	0.05	&	0.03	&	0.06	&	0.04			&0.16	&-0.07&	0.07&	0.83&	0.19\\
	CLASH-CXO		&	C	&	7	&	$\log M_{500}^\mathrm{HE}$	&	-0.06&	0.05	&	0.78	&	0.13	&	0.05	&	0.03	&	\multicolumn{2}{c}{[0]}	&--		&-0.07&	0.05&	0.88&	0.13\\
	\hline
	L13				&	F	&	29	&	$\log M_{500}$				&	-0.09&	0.03	&	1.44	&	0.27	&	0.07	&	0.03	&	0.06	&	0.02			&$-0.33	$&-0.09&	0.03&	1.59&	0.27\\
	L13				&	F	&	29	&	$\log M_{500}^\mathrm{HE}$	&	-0.09&	0.03	&	1.28	&	0.22	&	0.10	&	0.03	&	\multicolumn{2}{c}{[0]}	&--		&-0.09&	0.03&	1.51&	0.24\\
	L13				&	C	&	21	&	$\log M_{500}$				&	-0.08&	0.03	&	1.52	&	0.31	&	0.07	&	0.03	&	0.05	&	0.02			&$-0.22$	&-0.08&	0.04&	1.69&	0.34\\
	L13				&	C	&	21	&	$\log M_{500}^\mathrm{HE}$	&	-0.08&	0.03	&	1.36	&	0.25	&	0.09	&	0.03	&	\multicolumn{2}{c}{[0]}	&--		&-0.08&	0.03&	1.62&	0.28\\
	\hline
	CCCP-HE			&	F	&	33	&	$\log M_{500}$				&	-0.11&	0.04	&	1.16	&	0.36	&	0.12	&	0.06	&	0.11	&	0.05			&$-0.62$	&-0.12&	0.05&	1.43&	0.34\\
	CCCP-HE			&	F	&	33	&	$\log M_{500}^\mathrm{HE}$	&	-0.09&	0.03	&	0.83	&	0.17	&	0.17	&	0.03	&	\multicolumn{2}{c}{[0]}	&--		&-0.11&	0.04&	1.20&	0.25\\
	CCCP-HE			&	C	&	19	&	$\log M_{500}$				&	-0.02&	0.04	&	0.80	&	0.23	&	0.08	&	0.04	&	0.10	&	0.05			&$-0.48$	&-0.03&	0.04&	0.91&	0.25\\
	CCCP-HE			&	C	&	19	&	$\log M_{500}^\mathrm{HE}$	&	-0.01&	0.03	&	0.62	&	0.14	&	0.11	&	0.02	&	\multicolumn{2}{c}{[0]}	&--		&-0.02&	0.03&	0.76&	0.17\\
	\hline
	\end{tabular}
\end{table*}

\begin{figure*}
\begin{tabular}{cc}
\includegraphics[width=8.5cm]{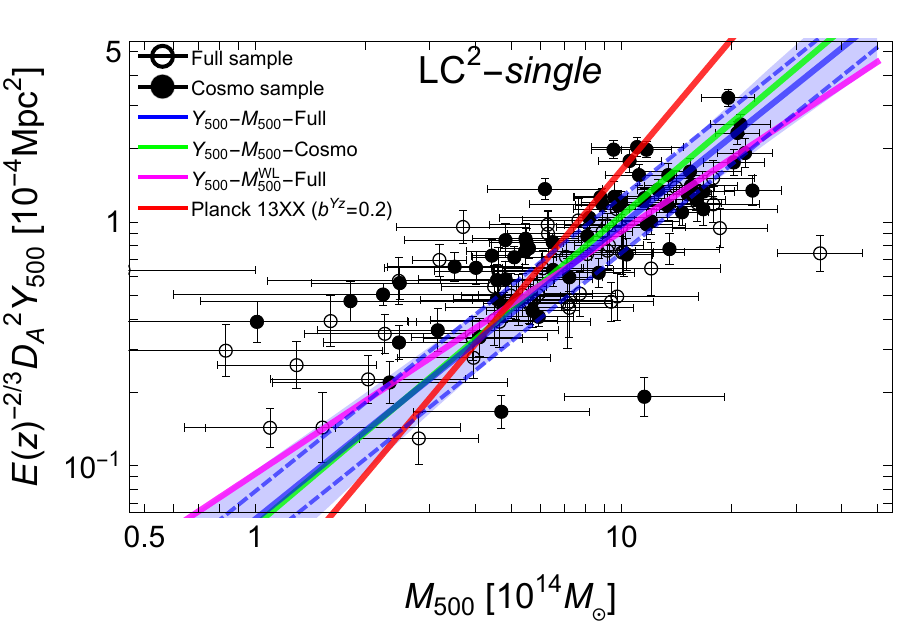}	&\includegraphics[width=8.5cm]{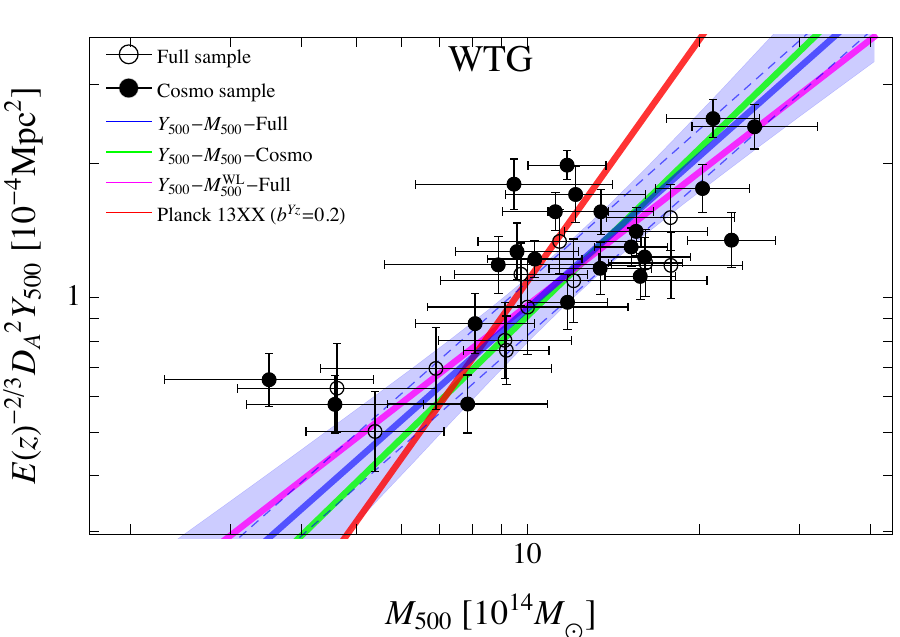}\\
\noalign{\smallskip}  
\includegraphics[width=8.5cm]{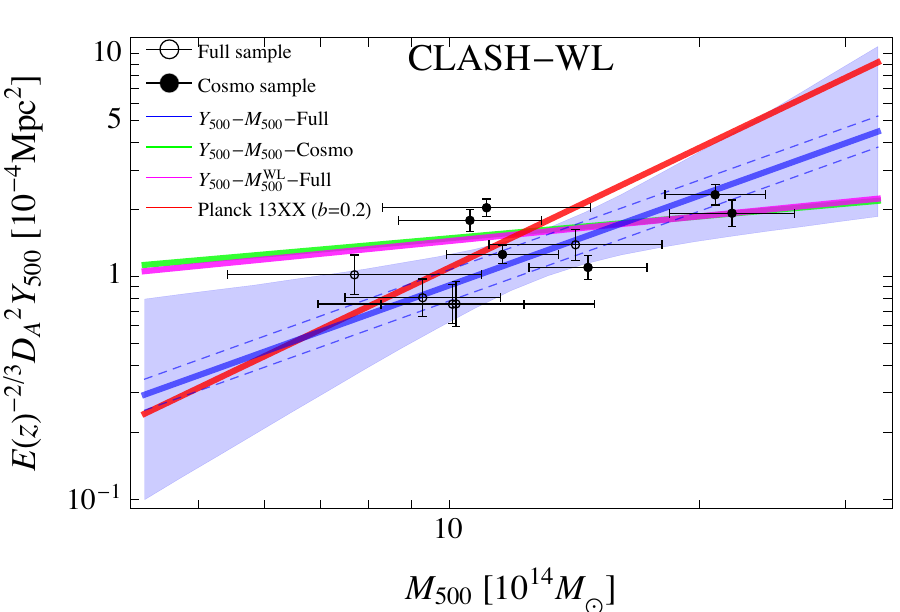} & \includegraphics[width=8.6cm]{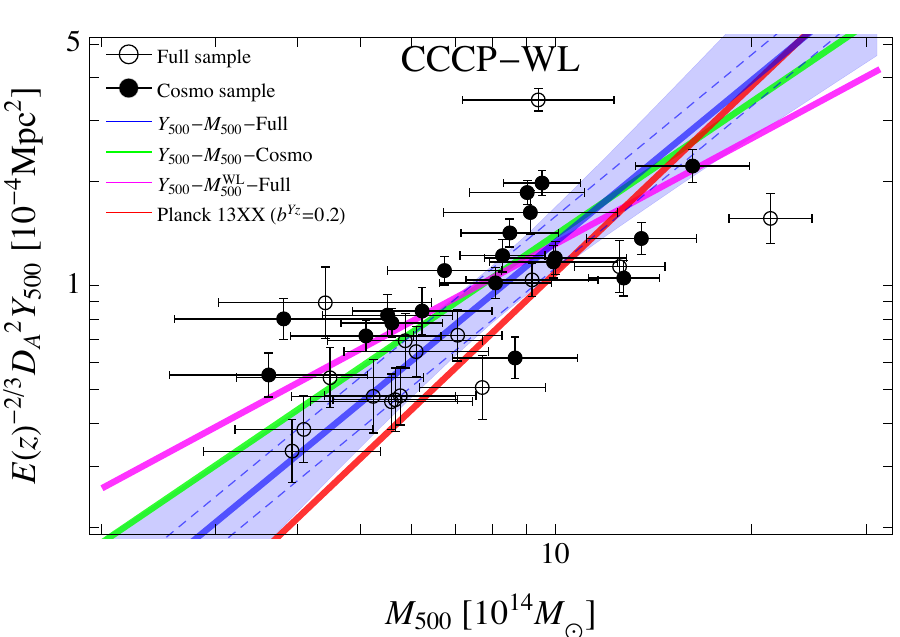} 
\end{tabular}
\caption{$M_{500}$ versus the spherically integrated Compton parameter for the WL samples. Filled points denote clusters in the cosmological sample; empty circles denote points in the full sample not included in the cosmological sample. The solid blue (green) line marks the mean fitted conditional scaling $Y_{500}$-$M_{500}$ for the full (cosmological) sample, while the dashed blue lines show this mean plus or minus the intrinsic scatter $\sigma_Y$. The shaded blue region encloses the 1-$\sigma$ confidence region. The magenta line plots the fitted conditional  $Y_{500}$-$M_{500}^\mathrm{WL}$ for the full SZ sample. The full red line plots the relation determined in \citet{planck_2013_XX} with bias $b^{Y_z}=0.2$. The top left, top right, bottom left, and bottom right panels represent the scaling relations calibrated with the LC$^2$-{\it single}, the WTG,  the CLASH-WL, and the CCCP-WL samples, respectively.}
\label{fig_M500_YSZ_WL}
\end{figure*}

\begin{figure}
       \resizebox{\hsize}{!}{\includegraphics{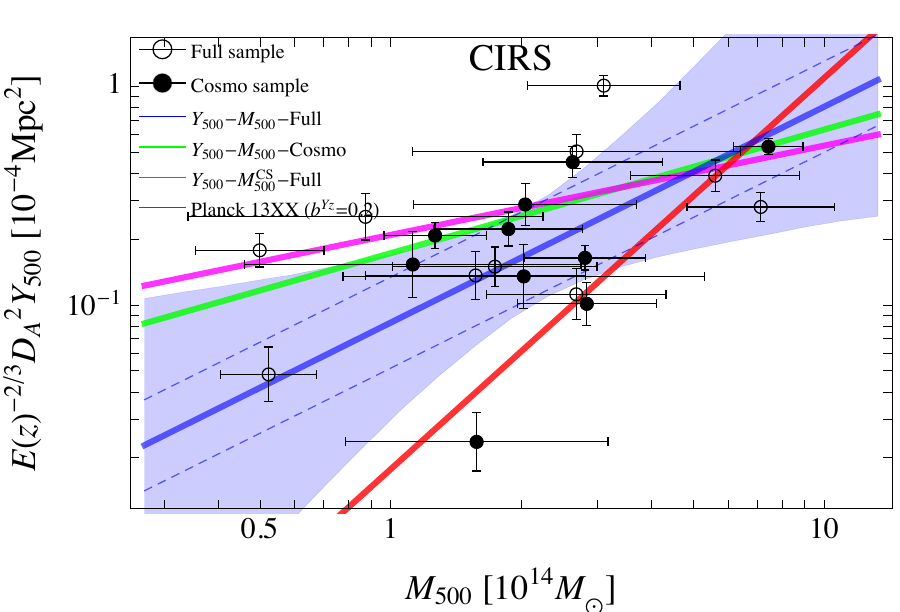}}
       \caption{$M_{500}$ versus the spherically integrated Compton parameter for the CIRS sample. The mass proxy is $M_{500}^\mathrm{CS}$. Points and lines are as in Fig.~\ref{fig_M500_YSZ_WL}.}
	\label{fig_M500_YSZ_CS}
\end{figure}

\begin{figure*}
\begin{tabular}{cc}
\includegraphics[width=8.5cm]{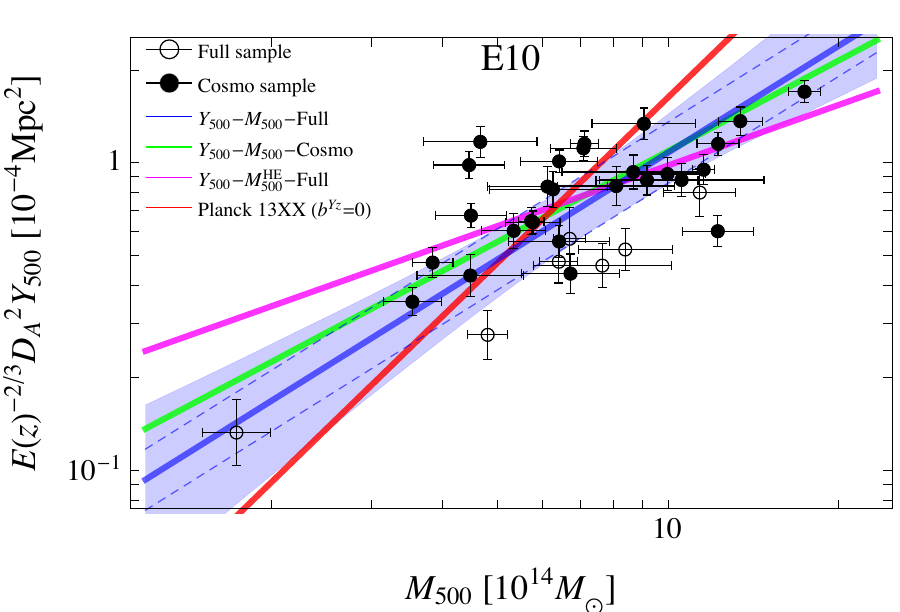} &\includegraphics[width=8.5cm]{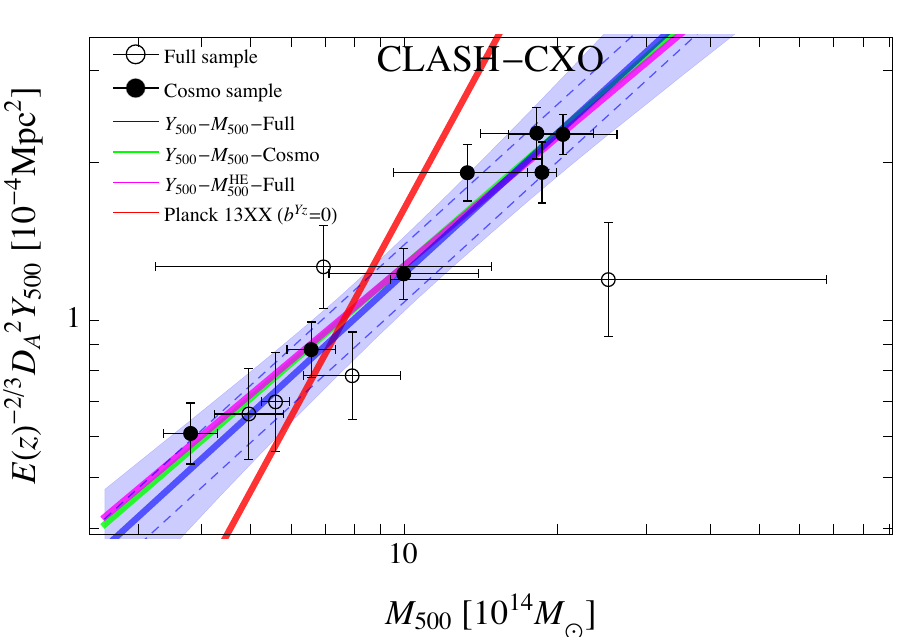}\\
\noalign{\smallskip}  
\includegraphics[width=8.5cm]{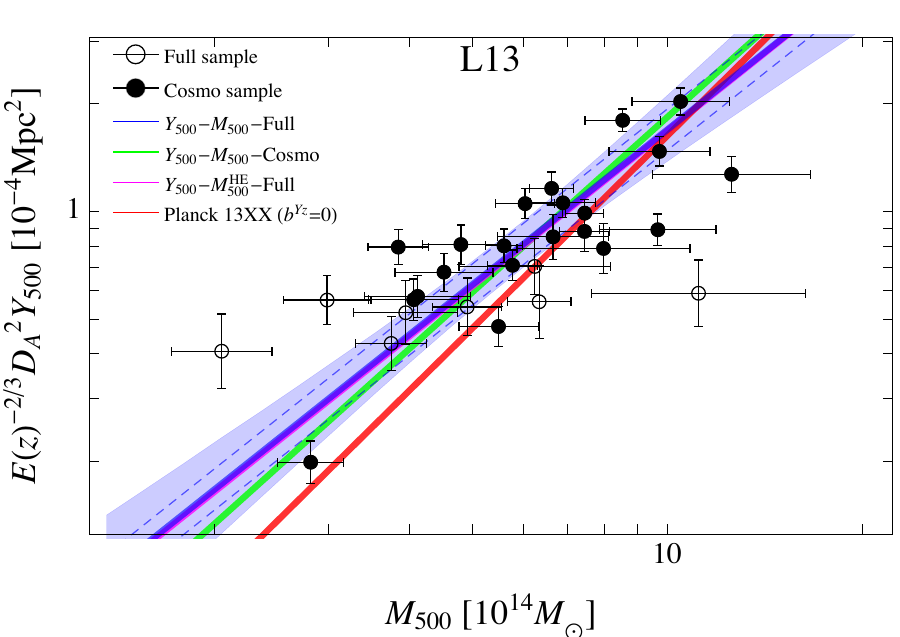} & \includegraphics[width=8.6cm]{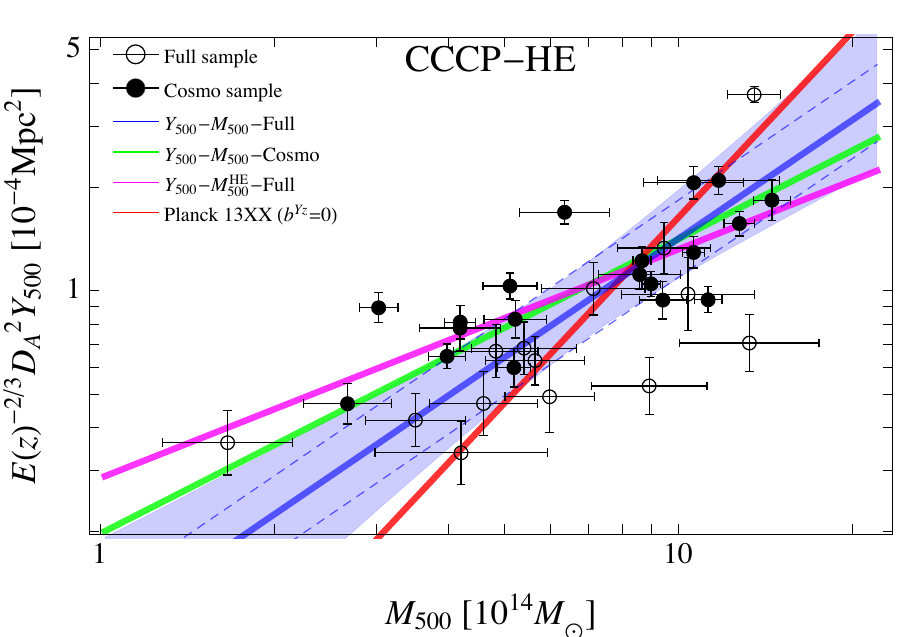}
\end{tabular}
\caption{$M_{500}$ versus the spherically integrated Compton parameter for the X-ray samples. The mass proxy is the HE mass. Points and lines are as in Fig.~\ref{fig_M500_YSZ_WL}. The bias for the scaling in \citet{planck_2013_XX} was fixed to $b^{Y_z}=0$. The top left, top right, bottom left, and bottom right panels represent the scaling relations calibrated with the E10, the CLASH-CXO, the L13 and the CCCP-HE samples, respectively.}
\label{fig_M500_YSZ_HE}
\end{figure*}

Regression results are summarised in Tables~\ref{tab_M500_YSZ_WL}, \ref{tab_M500_YSZ_CS}, and \ref{tab_M500_YSZ_HE}. We report results for both the conditional and the symmetric scaling relations. The intercept of the scaling relations were measured assuming that the proxy mass, i.e., either the WL, CS, or HE mass, is an unbiased tracer of the true mass. Any bias would consequently affect the normalisation of the $Y_{500}$-$M_{500}$ relation, which is known but for the bias between $M_{500}^\mathrm{Tr}$ and $M_{500}^\mathrm{Pr}$.

Slopes and scatters from WL, CS, or X-ray samples are in good agreement and results between the cosmological subsamples of clusters and the full ones are consistent. We will then focus mainly on WL clusters, which are more numerous and provide more accurate estimates.

\subsection{$Y_{500}$-$M_{500}$}

Regression results for the calibration sample based on WL masses are listed in Table~\ref{tab_M500_YSZ_WL} and plotted in Figure~\ref{fig_M500_YSZ_WL}. Results are consistent among the different samples. Estimates based on CLASH-WL are in agreement with other data-sets but, due to the small size of the sample, are affected by very large uncertainties and they will not be discussed in the following. The best estimates of the symmetric slope $\beta_{Y\textrm{-}Z}$ of the $Y_{500}$-$M_{500}^\mathrm{Tr}$ relation range from 1.4 to 1.9. These $\beta$ values are consistent within the errors with the self-similar slope of 5/3 and the fitting results of \citet[ table A.1]{planck_2013_XX} that lie around $\sim$1.6-1.8 ($1.79 \pm 0.06$ for the Cosmo sample).

As expected, the conditional scaling relation is flatter than the symmetric one. We found $\beta_{Y|Z} \sim 1.1$-$1.6$. These slopes are smaller than the reference values of 1.79 adopted to derive the $M^{Y_z}$ masses by the Planck team \citep{planck_2013_XX} . This discrepancy contributes to the mass dependent bias discussed in Section~\ref{sec_mass_bias}. The estimated intrinsic scatter on the WL mass is of order of 15-20 per cent, as expected from numerical simulations and as measured in \citetalias{se+et14_comalit_I}.

The intrinsic scatter $\sigma_{Y|Z}$ is of order of 15-30 per cent, in agreement with the estimate of $\sim$15 per cent in \citet{planck_2013_XX}.\footnote{We caution that we estimated the scatter along the vertical axis, whereas the quoted scatter in \citet{planck_2013_XX} was orthogonal to the regression.} As expected, we found a larger scatter in the LC$^2$-{\it single}, which we attribute to heterogeneity. 

\subsection{$Y_{500}$-$M^\mathrm{WL}_{500}$}

If we neglect the scatter in the mass proxy, we study the scaling between the SZ signal and the WL mass rather than the true mass. Due to the intrinsic scatter among the WL and the true mass, the scaling relation $Y_{500}$-$M_{500}^\mathrm{WL}$ is shallower then $Y_{500}$-$M_{500}^\mathrm{Tr}$ (see \citetalias{se+et14_comalit_I}) with slopes smaller by $\Delta \beta_{Y\textrm{-}Z} \sim0.1$-$0.2$. This is mainly due to clusters with large mass. Intrinsic scatter and observational uncertainties broaden the distribution. Massive clusters with the larger signal to noise ratio have the smaller relative uncertainty. At the low mass tail of the selected cluster mass function, the distribution is then mostly broadened by measurement errors (Eddington bias), whereas at large masses the intrinsic scatter plays a larger role.

The second main difference between the scalings is that the intrinsic scatter $\sigma_{Y|Z}$ in the conditional  probability $p(Y_{500}|M^\mathrm{WL}_{500})$ is larger than in $p(Y_{500}|M_{500})$. The SZ flux is a better proxy of the true mass rather than the WL mass and this translates in an inflated scatter in the scaling $Y_{500}$-$M^\mathrm{WL}_{500}$. In other words, when we measure the scatter of the $Y_{500}$-$M_{500}$, if we neglect the intrinsic scatter in the measured mass, we over-estimate the scatter of the SZ proxy and we under-estimate the statistical uncertainties on the scaling parameters.

\subsection{CS and X-ray samples}

Regression results for the CIRS sample (see Table~\ref{tab_M500_YSZ_CS} and Figure~\ref{fig_M500_YSZ_CS}), have too large uncertainties to draw firm conclusions.

The scaling relations derived from the samples with HE masses are similar to the WL case, see Table~\ref{tab_M500_YSZ_HE} and Figure~\ref{fig_M500_YSZ_HE}. The typical values of the symmetric slope are in the range $1.4 \ls \beta_{Y\textrm{-}Z} \ls 1.6$. Within the statistical uncertainties, the results are compatible with the self-similar scaling. The only exception is the relation derived for the CLASH-CXO sample, which is considerably flatter. However, this is the less numerous sample and few outliers may strongly affect the estimate.

The intrinsic scatter in the HE mass is estimated to be $\sim$15-25 per cent. Differently from numerical predictions, the scatter in HE masses is slightly larger than in WL masses (see \citetalias{se+et14_comalit_I}). The scatter in the SZ flux is of order of $\sim$10-30 per cent. Analogously to the WL case, we found that the $Y_{500}$-$M^\mathrm{HE}_{500}$ is shallower than the $Y_{500}$-$M_{500}$ relation.

Regardless of the calibration sample, we found results that are qualitatively and quantitatively in agreement. The CS case is not constraining, whereas derived scalings based on either WL or HE masses are fully consistent. On the other hand, the level of bias, i.e., the measured intercept of the relation, strongly depends on the assumed sample.

\subsection{Systematics}

Planck estimates of the flux $Y_{500}$ are obtained by linearly rescaling the flux $Y_{5r_{500}}$ determined within $r=5\times r_{500}$. The underlying assumption of a universal pressure profile with a fixed concentration \citep{planck_2013_XXIX} may underestimate the uncertainty on $Y_{500}$, which on turn may overestimate the estimate of the intrinsic scatter $\sigma_{Y|Z}$. A scatter of order of 10 (20) per cent in concentration induce an uncertainty of the order of $\sim$7 (15) per cent in the linear rescaling, which is smaller of (comparable to) the typical statistical uncertainty on $Y_{500}$ (usually of the order of $\sim15$ per cent). 

The pressure profiles were found to be distinctly more regular and to present less dispersion in the core than density profiles \citep{arn+al10}. The dispersion between $0.2r_{500}$ and $r_{500}$ is smaller than 20 per cent  \citep{arn+al10}. As far as the uncertainty on $Y_{500}$ due to instrumental accuracy is much larger than systematic errors due to modelling, the effect of linear rescaling is negligible. A proper assessment of this effect requires the study of the pressure profile in a sample of clusters alike the PSZ1 clusters.

\subsection{Correlated scatters}

From the statistical point of view, the estimates of the intrinsic scatters of the mass proxy and of the Compton flux with respect to the true mass are independent to first approximation. In fact, they spread the linear relation in orthogonal directions, and their effects should be disentangled with adequate data-sets. However, both scatters spread the relation, which can translates to some degree in a statistical anti-correlation. 

From the physical point of view, the picture is more complicated. Results from numerical simulations strongly depend on the adopted scheme for the gas physics.  \citet{sta+al10} found that the SZ flux is poorly correlated with the dark matter velocity dispersion but it is strongly correlated with other gas observables, e.g., X-ray bolometric luminosity, temperature, and gas mass fraction.

Triaxility and projection effects strongly impact WL estimates. Masses and concentrations of prolate clusters aligned with the line of sight can be significantly over-estimated. With such geometrical configuration, which is quite common in flux selected sample of clusters, the hydrostatic mass is affected too, since the projected scale radius is under-estimated and the HE mass over-estimated. However, the gas distribution is expected to be rounder than the matter distribution in clusters near to the equilibrium. Furthermore, the HE mass is not affected by the overall normalisation of the gas profile. The effect of triaxiality is then much more pronounced in the WL estimate. The measured projected Compton parameter $Y_\mathrm{cyl}$ is inflated too, with a consequent over-estimation of the spherically integrated flux $Y_{500}$. As for the HE mass, the effect is reduced by the rounder distribution of the gas.

A second major source of scatter is due to substructures. The related error in the WL mass depends on both substructure positions and method of analysis. If the shear is measured tangentially, massive clumps just outside the virial radius of the cluster can severely bias low the WL mass \citep{ras+al12}. Different configurations can bias the estimates in other directions. The measured projected mass is accurate at the level of $\sim10$ per cent for those clusters without any massive substructures nearby.

Matter substructures may be displaced by gas clumps, especially for merging clusters. Together with the dependence of the scatter on the specific measurement method, this makes very difficult to quantify the correlation among the scatter in the WL mass, the HE mass, or the SZ flux.

We found that the estimate of the intrinsic scatter between true mass and proxy mass (either WL, CS, or HE mass) and the estimate of the intrinsic scatter between true mass and SZ flux are usually slightly anti-correlated, see Tables~\ref{tab_M500_YSZ_WL}, \ref{tab_M500_YSZ_CS}, and \ref{tab_M500_YSZ_HE}. The degree of anti-correlation is larger in heterogeneous samples, i.e., the LC$^2$ sample. In fact, the marginalised probability distribution $p(\sigma_{X|Z},\sigma_{Y|Z})$ is quite round in the centre and slightly elongated at the tails, which increases the measured anti-correlation, see \citetalias{se+et14_comalit_I} and  \citetalias{se+et14_comalit_IV}.

This anti-correlation should then just reflect the intrinsic degeneracy of the statistical regression method rather than an intrinsic physical property, whose analysis we will address in a future paper.

\subsection{Mixture of Gaussians}

\begin{table}
\caption{Parameters of the scaling relation $Y_{500}$-$M_{500}$ of the full LC$^2$-{\it single} sample for different modelling of the intrinsic mass function. Col. 1: Parameter of the conditional scaling relation, see Table~\ref{tab_M500_YSZ_WL}. The log-mass function was approximated either with a single Gaussian (col.~2, $n_\mathrm{mix}=1$), or a mixture of two (col.~3, $n_\mathrm{mix}=2$), or three (col.~4, $n_\mathrm{mix}=3$) Gaussian functions. Listed values of the scatters refer to the logarithm to base 10. Quoted values are bi-weight estimators of the posterior probability distribution.}
\label{tab_M500_YSZ_LC2_mixture}
\centering
\begin{tabular}[c]{l  r@{$\,\pm\,$}lr@{$\,\pm\,$}lr@{$\,\pm\,$}l}
        \hline
        \noalign{\smallskip}
       Parameter	& \multicolumn{6}{c}{$n_\mathrm{mix}$}  \\
       	& \multicolumn{2}{c}{$1$} 	& \multicolumn{2}{c}{$2$} & \multicolumn{2}{c}{$3$}  \\
	\hline
	$\alpha_{Y|Z}$&		-0.28&	0.03	&	-0.28&	0.03&	-0.28&	0.03\\
	$\beta_{Y|Z}$&			1.22	&	0.20	&	1.20&	0.20&	1.22&	0.20\\
	$\sigma_{Y|Z}$&		0.11	&	0.05	&	0.12&	0.04&	0.11&	0.05\\
	$\sigma_{X|Z}$&		0.10	&	0.04&	0.09&	0.04&	0.10&	0.04\\
	\hline
	\end{tabular}
\end{table}

The mass function of galaxy clusters can be approximated with a Gaussian function in a large number of cases and selection functions, see \citetalias{se+et14_comalit_IV}. This simple modelling is very effective as far as the distribution of the independent variable is fairly unimodal \citep{kel07}. In fact, the Gaussian function provide a good approximation even in sparse samples, see \citetalias{se+et14_comalit_IV}.

We checked this working hypothesis. In Table~\ref{tab_M500_YSZ_LC2_mixture}, we report the results for the larger sample, i.e., the full LC$^2$-{\it single} sample, whose distribution of intrinsic true masses was approximated with a mixture of up to three Gaussian functions. Results are not statistically distinguishable and the simpler modelling should be preferred.

\section{Discussion}
\label{sec_disc}

We review how our results compare to previous works and theoretical expectations.

\subsection{Other works}

Previous works on the scaling relation between mass and integrated Compton parameter are discordant to some degree. Fair comparison is further complicated since slopes determined in different studies may actually refer to different statistical quantities, see Sec.~\ref{sec_slop}.

\citet{mar+al12} considered 18 galaxy clusters at $z\sim 0.2$ from the LoCuSS sample \citep[Local Cluster Substructure Survey,][]{oka+al10} observed with the Sunyaev-Zel'dovich Array. They found $\beta =  0.44\pm 0.12$ for the conditional $M_{500}^\mathrm{WL}$-$Y_{500}$ relation, with a 20 per cent intrinsic scatter. \citet{planck_int_III} considered 19 clusters mainly from the LoCuSS sample and found $\beta = 1.7 \pm 0.4$ for the orthogonal $Y_{500}$-$M_{500}^\mathrm{WL}$ relation. However, it was later understood that LoCuSS WL masses are biased low due to contamination effects and systematics in shape measurements \citep{oka+al13}. The underestimate of $M_{500}$ is of the order of 20 per cent \citep{oka+al13} and might be mass dependent.

\citet{roz+al14c} proposed a self-consistent method to derive scaling relations satisfying optical data from SDSS, X-ray data from ROSAT and Chandra, and SZ data from Planck. They derived a slope for the $Y_{500}$-$M_{500}$ relation of $1.71\pm 0.08$. The bias in the Planck masses can be estimated by equating the scaling relation in \citet{planck_2013_XX} to the relation in \citet{roz+al14c}. If we require that the predicted SZ fluxes are consistent at $M_{500}=6\times 10^{14}M_\odot$, we get $b\sim 0.26$.

Our results show some similarities with some other recent studies. \citet{gru+al14}, which performed WL analyses of 7 Planck clusters, found significant discrepancies between the weak lensing masses and the PSZ1 masses and a shallow slope, $\beta = 0.76 \pm 0.20$. They suggested that a size or redshift dependent bias could affect the analysis of the Planck clusters.

Comparing the Planck masses to the WL masses of the WTG clusters, \citet{lin+al14} found evidence for a mass dependence of the calibration ratio. They argued that the origin may hinge on systematic uncertainties in the X-ray temperature measurements used to calibrate the Planck cluster masses.

\subsection{Theoretical predictions}

Self-similar scaling relations aim at describing the intrinsic nature of a complete population of  galaxy clusters. SZ detected clusters are usually very massive haloes where evolution is mainly driven by gravitational processes and the self-similar behaviour should be retrieved. We found that the slope of the $Y_{500}$-$M_{500}$ relation is compatible with the self-similar prediction. 

The Planck cluster candidates were identified with some criteria that may separate the selected clusters from the undifferentiated population which self-similar models are based on. The very massive clusters in the Planck catalogue might be peculiar objects. Planck detected clusters are characterised by their global properties but we know little about their dynamical status. Irregular morphologies or ongoing merging events might impact the estimate of the cluster mass and the SZ flux to a different degree. In principle, WL estimates can give unbiased estimates even in complex systems, whereas clumpiness and irregularities play a bigger role in the intra-cluster medium distribution \citep{ron+al13}.

The roles of projection effects and orientation in a SZ selected sample should be better understood too. Clusters elongated towards the observer are more luminous and preferentially included in flux-limited samples. A basic spherical analysis over-estimates the lensing mass and the concentration of clusters elongated along the line of sight \citep{se+um11,gio+al14}. 

SZ estimates are affected too. The analysis of the visibility map constrains the projected flux $Y_\mathrm{cl}$, whereas the value of the spherically integrated $Y_{500}$ is usually derived by deprojection assuming spherical symmetry. However, the distribution of the intra-cluster medium is usually more spherical than the total matter so that triaxiality and orientation have a smaller impact in the estimate of the spherically integrated SZ flux \citep{ser+al13}. As a consequence, clusters elongated along the line of sight can make the scaling relation shallower at large WL mass. The above effect may induce a mass dependent intrinsic scatter on the WL or HE mass determination.

Even though the above effects can modify the $Y_{500}$-$M_{500}$ scaling, the statistical uncertainties we obtained are too large to ascertain their real impact.

\subsection{Number counts}

The slope and the normalisation of the scaling relation between $M_{500}$ and $Y_{500}$ strongly impact the expected number of observable clusters above a flux threshold. The Planck team argued that a large mass bias of the order of $b^{Y_z}\sim$ 40 per cent can reconcile their cosmological results from number counts with the analysis of the CMB \citep{planck_2013_XX}. This level of bias is in agreement with the mass calibrations based on the WTG and CLASH lensing samples, which implied $b^{Y_z}\sim$30-50 percent. On the other hand, the analysis of the CCCP-WL sample does not show such large bias.

The slope of the relation also impacts the number counts. A flatter scaling relation would determine a higher mass threshold for detection. This is the case of the slopes of the conditional scalings we measured ($1.2\ls \beta_{Y|Z}\ls1.6$), which are shallower than the slope of $\sim 1.8$ used in \citet{planck_2013_XX}.

Since massive clusters are rarer, the expected number count of massive clusters above a given flux threshold based on a flatter relation is then smaller than the prediction based on a steeper scaling. The flatter the relation, the larger the value of $\sigma_8$ required to match an observed number count. A steeper scaling may then bias low the measured amplitude of the power spectrum.

The above considerations reverse at low masses, where the flatter relation is above the steeper one in the $Y_{500}$-$M_{500}$ plane. At low masses, a steeper scaling biases high the estimated value of $\sigma_8$.

The effects of the slope at either low or large masses counterbalance each other to some degree. Furthermore, due to selection effects the mass of detectable clusters increases with redshift. A more quantitative assessment on the impact on the total number counts requires the knowledge of the completeness function of the Planck detections.

\section{Conclusions}
\label{sec_conc}

The effective use of scaling relations hinges on our ability to accurately measure cluster masses. WL and HE masses are accurate but scattered proxies. Numerical simulations have tried to ascertain the level of systematics affecting mass determinations \citep{ras+al12,nel+al14}. The bias in WL masses is mostly connected to irregular morphology and projection effects of the dark matter distribution \citep{men+al10,be+kr11}. The level of such bias is of order of 5-10 per cent with a large scatter of 10-25 per cent \citep{ras+al12,se+et14_comalit_I}. These are known effects which can be reduced with an optimal target selection.

Hydrostatic masses suffer from non-thermal sources of pressure in the intra-cluster medium and temperature inhomogeneity. The X-ray masses are biased low by a large amount  of 25-35 per cent with a sizeable scatter (see \citetalias{se+et14_comalit_I}). X-ray properties of galaxy clusters reported by competing groups reach discrepancies of 50 per cent in mass estimates \citep{roz+al14,se+et14_comalit_I}. 

Issues in instrumental calibration and methodological discrepancies in the data analysis may induce significant errors in the mass determination. Systematics differences in either WL or HE cluster mass impede a definite assessment of the mass bias and a robust calibration of the scaling relations. A consistent analysis of multi-wavelength observables from radio to optical to X-ray bands can help to single out the source of disagreement and to establish unbiased relations \citep{ser+al13,roz+al14c}.

Even though WL masses seem to be more accurate than X-ray estimates, the level of systematic uncertainty is still too high to accurately calibrate scaling relations. 

The SZ signal has been emerging as a very promising mass proxy. It should provide reliable mass estimates up to high redshifts even in irregular clusters \citep{sta+al10,bat+al12,kay+al12}. An accurate calibration is crucial for investigating the cluster physics and in the context of number counts of SZ detected clusters to constrain the cosmological parameters \citep{planck_2013_XX}. Notwithstanding the absolute normalisation, which is uncertain due to the systematic errors in measurements of WL or HE masses, we found trends between mass and SZ flux in notable agreement with theoretical predictions. The scaling is consistent with self-similarity and the intrinsic scatter is small. We showed that the conditional scaling $Y_{500}$-$M_{500}$ to be used in number count studies is shallower than the self-similar expectation due to the scattered nature of the relation.

The further step to obtain more accurate results is an improved characterisation of the selection function of the sample and the accurate calibration of the mass estimates.

The redshift evolution of scaling relation was not addressed in this paper. We adopted a self-similar evolution between mass and SZ flux \citep{gio+al13}. Due to selection limits, clusters at larger redshifts are on average more massive. Time evolution in the scaling might mimic mass dependent effects \citep{gru+al14,lin+al14}. The study of the evolution requires a very accurate knowledge of the cluster mass function and the selection function \citep{an+co14}. Ignoring them can led to biases larger than the quoted errors. The time-evolution of the scaling relation was addressed in \citetalias{se+et14_comalit_IV}.

\section*{Acknowledgements}
MS thanks Stefano Andreon for highlighting discussions. LM and MS acknowledge financial contributions from contracts ASI/INAF n.I/023/12/0 `Attivit\`a relative alla fase B2/C per la missione Euclid', PRIN MIUR 2010-2011 `The dark Universe and the cosmic evolution of baryons: from current surveys to Euclid', and PRIN INAF 2012 `The Universe in the box: multiscale simulations of cosmic structure'. SE acknowledges the financial contribution from contracts ASI-INAF I/009/10/0 and PRIN-INAF 2012 `A unique dataset to address the most compelling open questions about X-Ray Galaxy Clusters'.


\setlength{\bibhang}{2.0em}

\appendix

\section{Slopes}
\label{app_slop}

\begin{figure}
       \resizebox{\hsize}{!}{\includegraphics{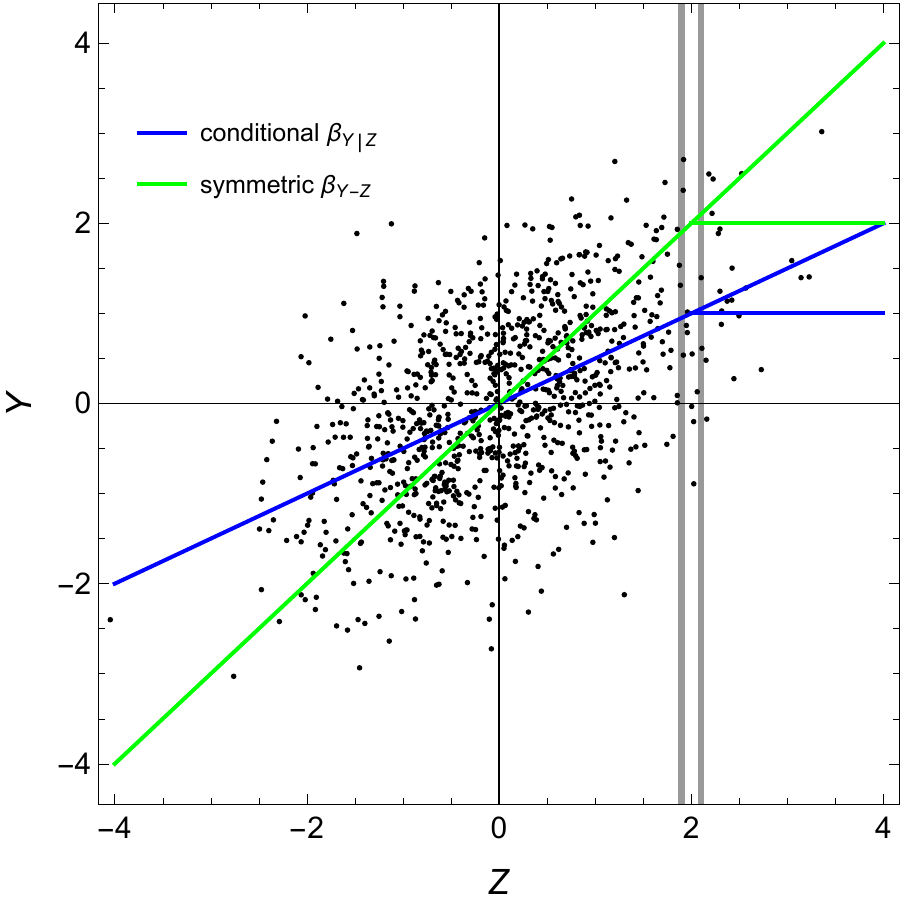}}
       \caption{
       The slopes of two different linear relations. The conditional scaling with slope $\beta_{Y|Z}$ (blue) tracks the most probable value of $Y$ given $Z$. The symmetric line with slope $\beta_{Y\textrm{-}Z}$ (green) follows the main direction. The points are drawn from a bivariate Gaussian with means $\mu_Z=\mu_Y=0$, standard deviations $\sigma_Z=\sigma_Y=1$ and correlation factor $\rho_{YZ}=0.5$.}
	\label{fig_slopes}
\end{figure}

The slope and the intercept of the `symmetric' scaling relation, Eq.~(\ref{eq_slop_1}), which better describes the joint relationship between two variables $Y$ and $Z$, can be related to the parameters of the `conditional' scaling relation, Eq.~(\ref{eq_slop_2}), which expresses the expected value of $Y$ given $Z$. In the `conditional' scaling relation, the symmetry between $Z$ and $Y$ is broken. As an illustrative example let us consider two variables, $Y$ and $Z$, whose probability distribution is a bivariate Gaussian centred in $\{ \mu_Y,\mu_Z\}$. The covariance matrix of the distribution has the variances $\sigma_Y^2$ and $\sigma_Z^2$ as diagonal elements, whereas the off-diagonal elements are given by $\rho_{YZ}~\sigma_Y\sigma_Z $, where $\rho_{YZ}$ is the correlation. In this simple case, if we want to predict $Y$ given $Z$ we know that the conditional probability $P(Y|Z)$ is a Gaussian distribution centred in
\beq
\label{eq_slop_3}
\mu_{Y|Z}=\mu_Y+\rho_{YZ} \frac{\sigma_Y}{\sigma_Z}(Z-\mu_Z),
\eeq
and with variance
\beq
\label{eq_slop_4}
\sigma_{Y|Z}^2=(1-\rho_{YZ}^2)\sigma_Y^2.
\eeq
By comparison between Eqs.~(\ref{eq_slop_2}, and \ref{eq_slop_4}), we can identify the slope of the conditional linear relation with
\beq
\label{eq_slop_5}
\beta_{Y|Z}=\rho_{YZ} \frac{\sigma_Y}{\sigma_Z},
\eeq
and the intercept with
\beq
\label{eq_slop_6}
\alpha_{Y|Z}=\mu_Y-\beta_{Y|Z}\mu_Z.
\eeq

If we want to study how $X$ and $Y$ co-evolve, we have to look to the principal direction of the distribution. This can be summarised by the line
\beq
\label{eq_slop_7}
\mu_{Y\textrm{-}Z}=\mu_Y+\beta_{Y\textrm{-}Z}(Z-\mu_Z),
\eeq
where the slope is given by the ratio of the eigenvalues of the covariance matrix
\beq
\label{eq_slop_8}
\beta_{Y\textrm{-}Z} =\frac{\sigma_Z^2-\sigma_Y^2-\sqrt{\sigma_Z^4+2(2\rho_{YZ}^2-1)\sigma_Z^2\sigma_Y^2+\sigma_Y^4}}{2\rho_{YZ}\ \sigma_Z \sigma_Y}.
\eeq
By comparing Eqs.~(\ref{eq_slop_1}, and \ref{eq_slop_8}), we can identify the intercept with
\beq
\label{eq_slop_9}
\alpha_{Y\textrm{-}Z}=\mu_Y-\beta_{Y\textrm{-}Z}\mu_Z.
\eeq
The two slopes are similar when the variable are highly correlated ($|\rho_{YZ}| \ls 1$) or when the dispersion in $Y$ is negligible with respect to $\sigma_Z$ and the relation is nearly flat. Otherwise, the slopes may significantly differ, see Fig.~\ref{fig_slopes}.

\end{document}